\documentclass[prd,reprint,onecolumn,notitlepage,preprintnumbers,superscriptaddress]{revtex4}
\usepackage{wrapfig}
\usepackage{amsfonts}
\usepackage{amsmath}
\usepackage{hyperref}
\usepackage{amssymb}
\usepackage[english]{babel}
\usepackage{graphicx}
\usepackage{epsfig}
\usepackage{bm}
\usepackage{longtable}
\usepackage{verbatim}
\usepackage{longtable}
\usepackage{color}
\usepackage{subfigure}
\usepackage[utf8]{inputenc}

\input{tcilatex}
\begin{document}

\title{Fermion and scalar phenomenology of a 2-Higgs doublet model with $S_3$
}
\author{A. E. C\'arcamo Hern\'andez}
\email{antonio.carcamo@usm.cl}
\affiliation{{\small Universidad T\'ecnica Federico Santa Mar\'{\i}a and Centro Cient%
\'{\i}fico-Tecnol\'ogico de Valpara\'{\i}so}\\
Casilla 110-V, Valpara\'{\i}so, Chile}
\author{I. de Medeiros Varzielas}
\email{ivo.de@soton.ac.uk}
\affiliation{{\small School of Physics and Astronomy, University of Southampton,}\\
Southampton, SO17 1BJ, U.K.}
\author{E. Schumacher}
\email{erik.schumacher@tu-dortmund.de}
\affiliation{{\small Fakult\"at f\"ur Physik, Technische Universit\"at Dortmund}\\
D-44221 Dortmund, Germany}

\begin{abstract}
We propose a 2-Higgs doublet model where the symmetry is extended by $%
S_{3}\otimes Z_{3}\otimes Z_{3}^{\prime }\otimes Z_{14}$ and the field
content is enlarged by extra $SU(2)_{L}$ singlet scalar fields. $S_3$ makes
the model predictive and leads to viable fermion masses and mixing. The
observed hierarchy of the quark masses arises from the $Z_{3}^{\prime }$ and 
$Z_{14}$ symmetries. The light neutrino masses are generated through a type
I seesaw mechanism with two heavy Majorana neutrinos. In the lepton sector
we obtain mixing angles that are nearly tri-bi-maximal, in an excellent
agreement with the observed lepton parameters. The vacuum expectation values
required for the model are naturally obtained from the scalar potential, and
we analyze the scalar sector properties further constraining the model
through rare top decays (like $t \to c h$), the $h \to \gamma \gamma$ decay channel and the $T$ and $S$ parameters.
\end{abstract}

\preprint{DO-TH 15/12}
\maketitle

\section{Introduction}

The flavor puzzle is not understood in the context of the Standard Model
(SM), which does not specify the Yukawa structures and has no justification
for the number of generations. As such, extensions addressing the fermion
masses and mixing are particularly appealing. With neutrino experiments
increasingly constraining the mixing angles in the leptonic sector many
models focus only on this sector, aiming to explain the near tri-bi-maximal
structure of the PMNS matrix through some non-Abelian symmetry.

Discrete flavor symmetries have shown a lot of promise and $S_3$, as the
smallest non-Abelian group has been considerably studied in the literature
since \cite{Pakvasa:1977in}, with interesting results for quarks, leptons or
both, and remains a popular group \cite{Cardenas:2012bg,Dias:2012bh,Dev:2012ns,Meloni:2012ci,Canales:2013cga,Ma:2013zca,Kajiyama:2013sza,Hernandez:2013hea,Hernandez:2014lpa,Hernandez:2014vta,Vien:2014vka,Ma:2014qra,Das:2014fea,Das:2015sca}%
. Other popular groups are the smallest groups with triplet representations,
particularly $A_4$ which has only a triplet and three distinct singlets. $%
A_4 $ was used in \cite{Ma:2001dn,Babu:2002dz,Altarelli:2005yp,Altarelli:2005yx,deMedeirosVarzielas:2005qg}
and more recently in \cite{Varzielas:2012ai,Ishimori:2012fg,Ahn:2013mva,Memenga:2013vc,Bhattacharya:2013mpa, Ferreira:2013oga,Felipe:2013vwa,Hernandez:2013dta,King:2013hj,Morisi:2013qna,Morisi:2013eca,Felipe:2013ie,Campos:2014lla,Hernandez:2015tna,Pramanick:2015qga}%
. With just triplets and singlet representations the groups $T_7$ \cite{Luhn:2007sy,Hagedorn:2008bc,Cao:2010mp,Luhn:2012bc,Kajiyama:2013lja,Bonilla:2014xla,Hernandez:2015cra,Arbelaez:2015toa}
and $\Delta(27)$ \cite{deMedeirosVarzielas:2006fc,Ma:2006ip,Varzielas:2012nn,Bhattacharyya:2012pi,Ma:2013xqa,Varzielas:2013sla,Aranda:2013gga,Varzielas:2013eta,Varzielas:2015aua}
are also promising as flavor symmetries. For recent reviews on the use of
discrete flavor groups, see Refs. \cite{King:2013eh,King:2014nza}.

In this work we make use of the $S_3$ group to formulate a 2-Higgs doublet
model (2HDM) with an extra $S_3\otimes Z_{3}\otimes Z_{3}^{\prime}\otimes
Z_{14}$ symmetry. Assigning the SM fermions under this symmetry and using
scalars transforming under the different irreducible representations of $S_3$, we provide an existence
proof of models leading to the viable mixing inspired quark textures presented in \cite{Hernandez:2014zsa}, by building a minimal realization.
We then consider the model in the lepton sector where
we obtain viable masses and mixing angles by using assignments that lead to
a charged lepton texture similar to that of the down-type quarks, with the
neutrino sector being completed through a type I seesaw. We discuss the
scalar potential in some detail, showing it leads to the Vacuum Expectation
Values (VEVs) used to obtain the fermion masses, and analyzing
phenomenological processes that constrain the parameters of the model such
as $t\to ch$ and $h \to \gamma \gamma$.

The paper is outlined as follows. In Section \ref{model} we describe the
field and symmetry content of the model, including a brief revision of the
quark mass and mixing angles presented in \cite{Hernandez:2014zsa} (Section %
\ref{Quarks}) and the equivalent analysis for the lepton sector (Section \ref%
{Leptons}). Section \ref{scalarpheno} contains the analysis of the
phenomenology associated with the extended scalar sector, presenting the
Yukawa couplings, an analysis of rare top decays, then considering the $h \to \gamma \gamma$ rate (Section \ref{gammagamma}) and the $T$ and $S$ parameters (Section \ref{TnS}). We present
our conclusions in Section \ref{Conclusions}. We relegate some technical
discussions that are relevant for the paper to the Appendix.

\section{The Model \label{model}}
We consider an extension of the SM with extra scalar fields and discrete symmetries, which reproduces the predictive mixing inspired textures proposed in Ref. \cite{Hernandez:2014zsa}, i.e. the Cabbibo mixing arises from the down-type quark sector whereas the up-type quark sector contributes to the remaining mixing angles. These textures describe the charged fermion masses and quark mixing pattern in terms of different powers of the Wolfenstein parameter $\lambda=0.225$ and order one parameters.
Because of the required mismatch between the down-type quark and up-type quark textures, to obtain these textures in a model we use two Higgs doublets distinguished by a symmetry (in our model, a $Z_3$). 
In the following, we describe our 2HDM with the inclusion of the $S_{3}\otimes Z_{3}\otimes Z_{3}^{\prime }\otimes Z_{14}$ discrete symmetry and four singlet scalar fields, assigned in a $S_3$ doublet, one $S_3$ trivial singlet and one $S_3$ non trivial singlet. We use the $S_3$ discrete group since it is the smallest non-Abelian group, having a doublet and two singlets as irreducible representations.
The full symmetry $\mathcal{G}$ of the model is broken spontaneously in two
steps: 
\begin{eqnarray}
&&\mathcal{G}=SU(3)_{C}\otimes SU(2)_{L}\otimes U\left( 1\right) _{Y}\otimes
S_{3}\otimes Z_{3}\otimes Z_{3}^{\prime }\otimes Z_{14}  \label{Group} \\
&&\hspace{35mm}\Downarrow \Lambda  \notag \\[3mm]
&&\hspace{15mm}SU\left( 3\right) _{C}\otimes SU\left( 2\right) _{L}\otimes
U\left( 1\right) _{Y}\otimes Z_{3}  \notag \\[3mm]
&&\hspace{35mm}\Downarrow \Lambda _{EW}  \notag \\[3mm]
&&\hspace{23mm}SU\left( 3\right) _{C}\otimes U\left( 1\right) _{em},  \notag
\end{eqnarray}%
where the different symmetry breaking scales satisfy the following hierarchy 
$\Lambda \gg \Lambda _{EW}$, where $\Lambda _{EW}=246$ GeV is the
electroweak symmetry breaking scale.

\begin{table}[tbp]
\centering%
\begin{tabular}{|c|ccc|cc|ccc|ccc|ccc|cc|}
\hline
Field & $q_{1L}$ & $q_{2L}$ & $q_{3L}$ & $U_{R}$ & $u_{3R}$ & $d_{1R}$ & $%
d_{2R}$ & $d_{3R}$ & $l_{1L}$ & $l_{2L}$ & $l_{3L}$ & $l_{1R}$ & $l_{2R}$ & $%
l_{3R}$ & $\nu _{1R}$ & $\nu _{2R}$ \\ \hline
$S_{3}$ & 1 & 1 & 1 & 2 & 1 & 1 & 1 & 1 & 1 & 1 & 1 & 1' & 1 & 1 & 1 & 1 \\ 
$Z_{3}$ & 0 & 0 & 1 & 0 & 1 & 2 & 2 & 1 & 2 & 0 & 0 & 1 & 0 & 0 & 0 & 0 \\ 
$Z_{3}^{\prime }$ & 0 & 0 & 0 & 0 & 0 & 0 & 0 & 0 & 0 & 0 & 0 & 2 & 0 & 0 & 0
& 0 \\ 
$Z_{14}$ & -3 & -2 & 0 & 1 & 0 & 4 & 3 & 3 & -3 & 0 & 0 & 4 & 5 & 3 & 0 & 0
\\ \hline
\end{tabular}%
\caption{Assignments of the SM fermions under the flavor symmetries. }
\label{ta:fermions}
\end{table}

\begin{table}[tbp]
\centering 
\begin{tabular}{|c|cc|c|cc|}
\hline
Field & $\phi _{1}$ & $\phi _{2}$ & $\xi $ & $\chi $ & $\zeta $ \\ \hline
$SU(2)_{L}$ & 2 & 2 & 1 & 1 & 1 \\ 
$S_{3}$ & 1 & 1 & 2 & 1 & 1' \\ 
$Z_{3}$ & 0 & 1 & 0 & 0 & 0 \\ 
$Z_{3}^{\prime }$ & 0 & 0 & 0 & 0 & 1 \\ 
$Z_{14}$ & 0 & 0 & 0 & -1 & 0 \\ \hline
\end{tabular}%
\caption{Assignments of the scalars under $SU(2)_L$ and the flavor
symmetries. }
\label{ta:scalars}
\end{table}

The content of the model, which includes the particle assignments under the
different symmetries, is shown in Tables \ref{ta:fermions} and \ref%
{ta:scalars}. The $S_{3}$ symmetry reduces the number of parameters in the
Yukawa sector of this 2HDM making it more predictive. The $Z_{3}$ symmetry
allows to completely decouple the bottom quark from the remaining down and
strange quarks. As can be seen from the scalar field assignments, the two
scalar $SU(2)_L$ doublets have different $Z_{3}$ charges ($\phi_1$ being
neutral). The $Z_{3}^{\prime }$ and $Z_{14}$ symmetries shape the
hierarchical structure of the quark mass matrices necessary to get a
realistic pattern of quark masses and mixing.

The Higgs doublets $\phi _{l}$ ($l=1,2$) acquire VEVs that break $SU(2)_{L}$
\begin{equation}
\phi _{l}=\left( 
\begin{array}{c}
0 \\ 
\frac{v_{l}}{\sqrt{2}}%
\end{array}%
\right) ,\hspace{1.5cm}l=1,2.
\end{equation}%
We decompose the Higgs fields around this minimum as 
\begin{equation}
\phi _{l}=\left( 
\begin{array}{c}
\varphi _{l}^{+} \\ 
\frac{1}{\sqrt{2}}\left( v_{l}+\rho _{l}+i\eta _{l}\right)%
\end{array}%
\right) =\left( 
\begin{array}{c}
\frac{1}{\sqrt{2}}\left( \omega _{l}+i\tau _{l}\right) \\ 
\frac{1}{\sqrt{2}}\left( v_{l}+\rho _{l}+i\eta _{l}\right)%
\end{array}%
\right) ,  \label{doublets}
\end{equation}%
where 
\begin{equation}
\left\langle \rho _{l}\right\rangle =\left\langle \eta _{l}\right\rangle
=\left\langle \omega _{l}\right\rangle =\left\langle \tau _{l}\right\rangle
=0,\hspace{2cm}\hspace{2cm}l=1,2.
\end{equation}

From an analysis of the scalar potential (see Appendix \ref{S3VEV}), we
obtain the following VEVs for the SM singlet scalars: 
\begin{equation}
\left\langle \xi \right\rangle =v_{\xi }\left( 1,0\right) ,\hspace{1cm}%
\left\langle \chi \right\rangle =v_{\chi },\hspace{1cm}\left\langle \zeta
\right\rangle =v_{\zeta },  \label{VEV}
\end{equation}%
i.e., the VEV of $\xi $ is aligned as $(1,0)$ in the $S_{3}$ direction.

For the up and down-type quarks, the Yukawa terms invariant under the
symmetries are 
\begin{equation}
\tciLaplace _{Y}^{U}=\varepsilon _{33}^{\left( u\right) }\overline{q}_{3L}%
\widetilde{\phi }_{1}u_{3R}+\varepsilon _{23}^{\left( u\right) }\overline{q}%
_{2L}\widetilde{\phi }_{2}u_{3R}\frac{\chi ^{2}}{\Lambda ^{2}}+\varepsilon
_{13}^{\left( u\right) }\overline{q}_{1L}\widetilde{\phi }_{2}u_{3R}\frac{%
\chi ^{3}}{\Lambda ^{3}}+\varepsilon _{22}^{\left( u\right) }\overline{q}%
_{2L}\widetilde{\phi }_{1}U_{R}\frac{\xi \chi ^{3}}{\Lambda ^{4}}%
+\varepsilon _{11}^{\left( u\right) }\overline{q}_{1L}\widetilde{\phi }%
_{1}U_{R}\frac{\xi \chi ^{4}\zeta ^{3}}{\Lambda ^{8}}+h.c.  \label{LU}
\end{equation}%
\begin{equation}
\tciLaplace _{Y}^{D}=\varepsilon _{33}^{\left( d\right) }\overline{q}%
_{3L}\phi _{1}d_{3R}\frac{\chi ^{3}}{\Lambda ^{3}}+\varepsilon _{22}^{\left(
d\right) }\overline{q}_{2L}\phi _{2}d_{2R}\frac{\chi ^{5}}{\Lambda ^{5}}%
+\varepsilon _{12}^{\left( d\right) }\overline{q}_{1L}\phi _{2}d_{2R}\frac{%
\chi ^{6}}{\Lambda ^{6}}+\varepsilon _{21}^{\left( d\right) }\overline{q}%
_{2L}\phi _{2}d_{1R}\frac{\chi ^{6}}{\Lambda ^{6}}+\varepsilon _{11}^{\left(
d\right) }\overline{q}_{1L}\phi _{2}d_{1R}\frac{\chi ^{7}}{\Lambda ^{7}}+h.c.
\label{LD}
\end{equation}%
The invariant Yukawa terms for charged leptons and neutrinos are 
\begin{equation}
\tciLaplace _{Y}^{l}=\varepsilon _{33}^{\left( l\right) }\overline{l}%
_{3L}\phi _{1}l_{3R}\frac{\chi ^{3}}{\Lambda ^{3}}+\varepsilon _{23}^{\left(
l\right) }\overline{l}_{2L}\phi _{1}l_{3R}\frac{\chi ^{3}}{\Lambda ^{3}}%
+\varepsilon _{22}^{\left( l\right) }\overline{l}_{2L}\phi _{1}l_{2R}\frac{%
\chi ^{5}}{\Lambda ^{5}}+\varepsilon _{32}^{\left( l\right) }\overline{l}%
_{3L}\phi _{1}l_{2R}\frac{\chi ^{5}}{\Lambda ^{5}}+\varepsilon _{11}^{\left(
l\right) }\overline{l}_{1L}\phi _{2}l_{1R}\frac{\chi ^{7}\zeta }{\Lambda ^{8}%
}+h.c.  \label{Ll}
\end{equation}%
\begin{eqnarray}
\tciLaplace _{Y}^{\nu } &=&\varepsilon _{11}^{\left( \nu \right) }\overline{l%
}_{1L}\widetilde{\phi }_{2}\nu _{1R}\frac{\chi ^{3}}{\Lambda ^{3}}%
+\varepsilon _{12}^{\left( \nu \right) }\overline{l}_{1L}\widetilde{\phi }%
_{2}\nu _{2R}\frac{\chi ^{3}}{\Lambda ^{3}}+\varepsilon _{21}^{\left( \nu
\right) }\overline{l}_{2L}\widetilde{\phi }_{1}\nu _{1R}+\varepsilon
_{22}^{\left( \nu \right) }\overline{l}_{2L}\widetilde{\phi }_{1}\nu _{2R}{%
+\varepsilon _{31}^{\left( \nu \right) }\overline{l}_{3L}\widetilde{\phi }%
_{1}\nu _{1R}+\varepsilon _{32}^{\left( \nu \right) }\overline{l}_{3L}%
\widetilde{\phi }_{1}\nu _{2R}}  \notag \\
&&{\ }+M_{1}\overline{\nu }_{1R}\nu _{1R}^{c}+M_{2}\overline{\nu }_{2R}\nu
_{2R}^{c}+M_{12}\overline{\nu }_{1R}\nu _{2R}^{c}+h.c.  \label{Lnu}
\end{eqnarray}
The $Z_{14}$ symmetry is the smallest cyclic symmetry that allows $\frac{%
\chi ^{7}}{\Lambda ^{7}}$ in the Yukawa terms responsible for the down quark
and electron masses, which we want to suppress by $\lambda ^{7}$ ($\lambda
=0.225$ is one of the Wolfenstein parameters) without requiring small
dimensionless Yukawa couplings. Furthermore, the $Z_{3}^{\prime }$ symmetry
is responsible for coupling the scalar $\zeta $ with $U_{R}$ as well as with 
$l_{1R}$, which helps to explain the smallness of the up quark and electron
mass in this model. The hierarchy of charged fermion masses and quark mixing
matrix elements is therefore explained by both the $Z_{3}^{\prime }$ and $%
Z_{14}$ symmetries. Given that in this scenario the quark masses are related
with the quark mixing parameters, we set the VEVs of the $SU(2)_{L}$ singlet
scalars with respect to the Wolfenstein parameter $\lambda $ and the new
physics scale $\Lambda $: 
\begin{equation}
v_{\xi }\sim v_{\zeta }\sim v_{\chi }=\lambda \Lambda .  \label{VEVsize}
\end{equation}%
These scalars therefore acquire VEVs at a scale unrelated with $\Lambda
_{EW} $. We have checked numerically that this regime is a valid minimum of
the global potential for a suitable region of the parameter space (see
Appendix \ref{S3VEV}). As we will see in the following sections, in order to
obtain realistic fermion masses and mixing without requiring a strong
hierarchy among the Yukawa couplings, the VEVs of the $SU(2)_{L}$ doublets ($%
v_{1}$ and $v_{2}$) should be of the same order of magnitude.

\subsection{Quark masses and mixing \label{Quarks}}

Using Eqs. (\ref{LU}) and (\ref{LD}) we find the mass matrices for up and
down-type quarks in the form: 
\begin{equation}
M_{U}=\frac{v}{\sqrt{2}}\left( 
\begin{array}{ccc}
c_{1}\lambda ^{8} & 0 & a_{1}\lambda ^{3} \\ 
0 & b_{1}\lambda ^{4} & a_{2}\lambda ^{2} \\ 
0 & 0 & a_{3}%
\end{array}%
\right) ,\hspace{1cm}\hspace{1cm}M_{D}=\frac{v}{\sqrt{2}}\left( 
\begin{array}{ccc}
e_{1}\lambda ^{7} & f_{1}\lambda ^{6} & 0 \\ 
e_{2}\lambda ^{6} & f_{2}\lambda ^{5} & 0 \\ 
0 & 0 & g_{1}\lambda ^{3}%
\end{array}%
\right) ,  \label{Quarktextures}
\end{equation}
where $a_{k}$ ($k=1,2,3$), $b_{1}$, $c_{1}$, $g_{1}$, $f_{1}$, $f_{2}$, $%
e_{1}$ and $e_2$ are $\mathcal{O}(1)$ parameters. Here we
assume that all dimensionless parameters given in Eq. (\ref{Quarktextures})
are real excepting $a_{3}$, which we assume to be complex. These are the
viable quark textures presented in \cite{Hernandez:2014zsa}, which we
briefly review here.

The hermitian combinations $M_{U}M_{U}^{\dagger }$ and $M_{D}M_{D}^{T}$ are 
\begin{eqnarray}
M_{U}M_{U}^{\dagger } &=&\allowbreak \frac{v^{2}}{2}\left( 
\begin{array}{ccc}
\left\vert a_{1}\right\vert ^{2}\lambda ^{6}+c_{1}^{2}\lambda ^{16} & 
a_{1}a_{2}\lambda ^{5} & a_{1}a_{3}\lambda ^{3} \\ 
a_{1}^{\ast }a_{2}\lambda ^{5} & a_{2}^{2}\lambda ^{4}+b_{1}^{2}\lambda ^{8}
& a_{2}a_{3}\lambda ^{2} \\ 
a_{1}^{\ast }a_{3}\lambda ^{3} & a_{2}a_{3}\lambda ^{2} & a_{3}^{2}%
\end{array}%
\right) ,  \label{Musquad} \\
M_{D}M_{D}^{T} &=&\allowbreak \frac{v^{2}}{2}\left( 
\begin{array}{ccc}
\lambda ^{14}e_{1}^{2}+\lambda ^{12}f_{1}^{2} & e_{1}e_{2}\lambda
^{13}+f_{1}f_{2}\lambda ^{11} & 0 \\ 
e_{1}e_{2}\lambda ^{13}+f_{1}f_{2}\lambda ^{11} & \lambda
^{12}e_{2}^{2}+\lambda ^{10}f_{2}^{2} & 0 \\ 
0 & 0 & \lambda ^{6}g_{1}^{2}%
\end{array}%
\right) ,
\end{eqnarray}%
and are approximately diagonalized by unitary rotation matrices $R_{U}$ and $%
R_{D}$: 
\begin{eqnarray}
R_{U}^{\dagger }M_{U}M_{U}^{\dagger }R_{U} &=&\left( 
\begin{array}{ccc}
m_{u}^{2} & 0 & 0 \\ 
0 & m_{c}^{2} & 0 \\ 
0 & 0 & m_{t}^{2}%
\end{array}%
\right) ,\hspace{1cm}\hspace{1cm}R_{U}\simeq \left( 
\begin{array}{ccc}
c_{13} & s_{13}s_{23}e^{i\delta } & -c_{23}s_{13}e^{i\delta } \\ 
0 & c_{23} & s_{23} \\ 
s_{13}e^{-i\delta } & -c_{13}s_{23} & c_{13}c_{23}%
\end{array}%
\right) \allowbreak ,  \label{quarktrafo} \\
R_{D}^{T}M_{D}M_{D}^{T}R_{D} &=&\left( 
\begin{array}{ccc}
m_{d}^{2} & 0 & 0 \\ 
0 & m_{s}^{2} & 0 \\ 
0 & 0 & m_{b}^{2}%
\end{array}%
\right) ,\hspace{1cm}\hspace{1cm}R_{D}=\left( 
\begin{array}{ccc}
c_{12} & s_{12} & 0 \\ 
-s_{12} & c_{12} & 0 \\ 
0 & 0 & 1%
\end{array}%
\right) ,\allowbreak
\end{eqnarray}%
where $c_{ij}=\cos \theta _{ij}$, $s_{ij}=\sin \theta _{ij}$ (with $i\neq j$
and$\ i,j=1,2,3$). $\theta _{ij}$ and $\delta $ are the quark mixing angles
and the CP violating phase, respectively, in the usual parametrization. They
are given by 
\begin{eqnarray}
\tan \theta _{12} &\simeq &\frac{f_{1}}{f_{2}}\lambda ,\hspace{1cm}\tan
\theta _{23}\simeq \frac{a_{2}}{a_{3}}\lambda ^{2},  \label{quarkangles} \\
\tan \theta _{13} &\simeq &\frac{\left\vert a_{1}\right\vert }{a_{3}}\lambda
^{3},\hspace{1cm}\delta =-\arg \left( a_{1}\right) .  \notag
\end{eqnarray}

Therefore, the up and down-type quark masses are approximately given by 
\begin{eqnarray}
m_{u} &\simeq &c_{1}\lambda ^{8}\frac{v}{\sqrt{2}},\hspace{1cm}\hspace{1cm}%
m_{c}\simeq b_{1}\lambda ^{4}\frac{v}{\sqrt{2}},\hspace{1cm}\hspace{1cm}%
m_{t}\simeq a_{3}\frac{v}{\sqrt{2}},  \label{Utypequarkmasses} \\
m_{d} &\simeq &\left\vert e_{1}f_{2}-e_{2}f_{1}\right\vert \frac{\lambda ^{7}%
}{\sqrt{2}}v,\hspace{1cm}\hspace{1cm}m_{s}\simeq f_{2}\lambda ^{5}\frac{v}{%
\sqrt{2}},\hspace{1cm}\hspace{1cm}m_{b}\simeq g_{1}\lambda ^{3}\frac{v}{%
\sqrt{2}}.  \label{Dquarkmasses}
\end{eqnarray}

We also find that the CKM quark mixing matrix is approximately 
\begin{equation}
V_{CKM}=R_{U}^{\dagger }R_{D}\simeq \left( 
\begin{array}{ccc}
c_{12}c_{13} & c_{13}s_{12} & e^{i\delta }s_{13} \\ 
e^{-i\delta }c_{12}s_{13}s_{23}-c_{23}s_{12} & c_{12}c_{23}+e^{-i\delta
}s_{12}s_{13}s_{23} & -c_{13}s_{23} \\ 
-s_{12}s_{23}-e^{-i\delta }c_{12}c_{23}s_{13} & c_{12}s_{23}-e^{-i\delta
}c_{23}s_{12}s_{13} & c_{13}c_{23}%
\end{array}%
\right) \allowbreak .
\end{equation}

It is noteworthy that Eq. (\ref{Quarktextures}) provides an elegant
understanding of all SM fermion masses and mixing angles through their
scalings by powers of the Wolfenstein parameter $\lambda =0.225$ with $%
\mathcal{O}(1)$ coefficients.

The Wolfenstein parametrization \cite{Wolfenstein:1983yz} of the CKM matrix
is: 
\begin{equation}
V_{W}\simeq \left( 
\begin{array}{ccc}
1-\frac{\lambda ^{2}}{2} & \lambda & A\lambda ^{3}(\rho -i\eta ) \\ 
-\lambda & 1-\frac{\lambda ^{2}}{2} & A\lambda ^{2} \\ 
A\lambda ^{3}(1-\rho -i\eta ) & -A\lambda ^{2} & 1%
\end{array}%
\right) ,  \label{wolf}
\end{equation}%
with 
\begin{eqnarray}
\lambda &=&0.22537\pm 0.00061,\quad \quad \quad A=0.814_{-0.024}^{+0.023}, \\
\quad \overline{{\rho }} &=&0.117\pm 0.021,\quad \quad \quad \overline{{\eta 
}}=0.353\pm 0.013, \\
\overline{{\rho }} &\simeq &\rho \left( 1-\frac{{\lambda }^{2}}{2}\right)
,\quad \quad \quad \overline{{\eta }}\simeq \eta \left( 1-\frac{{\lambda }%
^{2}}{2}\right) .
\end{eqnarray}%
From the comparison with (\ref{wolf}), we find: 
\begin{eqnarray}
a_{3} &\simeq &1,\quad \quad \quad a_{2}\simeq A\simeq 0.81,\quad \quad
\quad a_{1}\simeq -A\sqrt{\rho ^{2}+\eta ^{2}}e^{i\delta }\simeq
-0.3e^{i\delta }, \\
\delta &=&67^{\circ },\quad \quad \quad b_{1}\simeq \frac{m_{c}}{\lambda
^{4}m_{t}}\simeq 1.43,\quad \quad \quad c_{1}\simeq \frac{m_{u}}{\lambda
^{8}m_{t}}\simeq 1.27.
\end{eqnarray}%
Note that $a_{1}$ is required to be complex, as previously assumed, and its
magnitude is a bit smaller than the remaining $\mathcal{O}(1)$ coefficients.

Since the charged fermion masses and quark mixing hierarchy arises from the $%
Z_{3}^{\prime }\otimes Z_{14}$ symmetry breaking, and in order to have the
right value of the Cabbibo mixing, we need $e_{2}\approx f_{2}$%
. We fit the parameters $e_{1}$, $f_{1}$, $f_{2}$ and $g_{1}$ in Eq. (\ref%
{Quarktextures}) to reproduce the down-type quark masses and quark mixing
parameters. As can be seen from the above formulas, the quark
sector of our model contains ten effective free parameters, i.e., $|a_1|$, $%
a_2$, $a_3$, $b_1$, $c_1$, $e_{1}$, $f_{1}$, $f_{2}$, $g_{1}$ and the phase $%
\arg(a_1)$, to describe the quark mass and mixing pattern, which is
characterized by ten physical observables, i.e., the six quark masses, the
three mixing angles and the CP violating phase. Furthermore, in our model these parameters are of
the same order of magnitude. The results for the down-type
quark masses, the three quark mixing angles and the CP violating phase $%
\delta $ in Tables \ref{Observables} and \ref{obs} correspond to the best
fit values:
\begin{equation}
e_{1}\simeq 0.84,\hspace{1cm}f_{1}\simeq 0.4,\hspace{1cm}f_{2}\simeq 0.57,%
\hspace{1cm}g_{1}\simeq 1.42.
\end{equation}
As pointed out in \cite{Hernandez:2014zsa}, the CKM matrix in
our model is consistent with the experimental data. The agreement of our
model with the experimental data is as good as in the models of Refs. \cite%
{Branco:2010tx,King:2013hj,Hernandez:2013hea,Hernandez:2014vta,CarcamoHernandez:2010im,Bhattacharyya:2012pi,Campos:2014lla}
and better than, 
for example, those in Refs.~\cite%
{Chen:2007afa,Xing:2010iu,Branco:2011wz,CarcamoHernandez:2012xy, Vien:2014ica,Abbas:2014ewa,Ishimori:2014jwa,Ishimori:2014nxa}.
The obtained and experimental values of the magnitudes of the CKM
parameters, i.e., three quark mixing parameters and the CP violating phase $%
\delta $ are shown in Table \ref{Observables}. The experimental values of
the CKM magnitudes and the Jarlskog invariant are taken from Ref. \cite%
{Agashe:2014kda}, whereas the experimental values of the quark
masses, which are given at the $M_{Z}$ scale, have been taken from Ref. \cite%
{Bora:2012tx}.

\begin{table}[tbh]
\begin{center}
\begin{tabular}{c|l|l}
\hline\hline
Observable & Model value & Experimental value \\ \hline
$m_{u}(MeV)$ & \quad $1.47$ & \quad $1.45_{-0.45}^{+0.56}$ \\ \hline
$m_{c}(MeV)$ & \quad $641$ & \quad $635\pm 86$ \\ \hline
$m_{t}(GeV)$ & \quad $172.2$ & \quad $172.1\pm 0.6\pm 0.9$ \\ \hline
$m_{d}(MeV)$ & \quad $3.00$ & \quad $2.9_{-0.4}^{+0.5}$ \\ \hline
$m_{s}(MeV)$ & \quad $59.2$ & \quad $57.7_{-15.7}^{+16.8}$ \\ \hline
$m_{b}(GeV)$ & \quad $2.82$ & \quad $2.82_{-0.04}^{+0.09}$ \\ \hline
\end{tabular}%
\end{center}
\caption{Model and experimental values of the quark masses.}
\label{Observables}
\end{table}

\begin{table}[tbh]
\begin{center}
\begin{tabular}{c|l|l|}
\hline\hline
Observable & Model value & Experimental value \\ \hline
$\sin \theta _{12}$ & \quad $0.2257$ & \quad $0.2254$ \\ \hline
$\sin \theta _{23}$ & \quad $0.0412$ & \quad $0.0413$ \\ \hline
$\sin \theta _{13}$ & \quad $0.00352$ & \quad $0.00350$ \\ \hline
$\delta $ & \quad $68^{\circ }$ & \quad $68^{\circ }$ \\ \hline\hline
\end{tabular}%
\end{center}
\caption{Model and experimental values of CKM parameters.}
\label{obs}
\end{table}

\subsection{Lepton masses and mixing \label{Leptons}}

This $S_{3}$ flavor model obtains the viable quark textures proposed in \cite%
{Hernandez:2014zsa} as shown in section \ref{Quarks}. We now proceed to
analyze the lepton sector of the model. From the charged lepton Yukawa terms
of Eq. (\ref{Ll}) it follows that the charged lepton mass matrix takes the
following form: 
\begin{equation}
M_{l}=\frac{v}{\sqrt{2}}\left( 
\begin{array}{ccc}
x_{1}\lambda ^{8} & 0 & 0 \\ 
0 & y_{1}\lambda ^{5} & z_{1}\lambda ^{3} \\ 
0 & y_{2}\lambda ^{5} & z_{2}\lambda ^{3}%
\end{array}%
\right) .  \label{Ml}
\end{equation}
where $x_{1}$, $y_{1}$, $y_{2}$, $z_{1}$, $z_{2}$, are $\mathcal{O}(1)$ parameters, assumed to be real, for simplicity.

Then, the charged lepton mass matrix satisfies the following relations:
\begin{equation}
M_{l}M_{l}^{T}=\frac{v^{2}}{2}\left( 
\begin{array}{ccc}
x_{1}^{2}\lambda ^{16} & 0 & 0 \\ 
0 & z_{1}^{2}\lambda ^{6}+y_{1}^{2}\lambda ^{10} & z_{1}z_{2}\lambda
^{6}+y_{1}y_{2}\lambda ^{10} \\ 
0 & z_{1}z_{2}\lambda ^{6}+y_{1}y_{2}\lambda ^{10} & z_{2}^{2}\lambda
^{6}+y_{2}^{2}\lambda ^{10}
\end{array}%
\right) ,
\end{equation}
\begin{equation}
M_{l}^{T}M_{l}=\frac{v^{2}}{2}\left( 
\begin{array}{ccc}
x_{1}^{2}\lambda ^{16} & 0 & 0 \\ 
0 & \left( y_{1}^{2}+y_{2}^{2}\right) \lambda ^{10} & \left(
y_{1}z_{1}+y_{2}z_{2}\right) \lambda ^{8} \\ 
0 & \left( y_{1}z_{1}+y_{2}z_{2}\right) \lambda ^{8} & \left(
z_{1}^{2}+z_{2}^{2}\right) \lambda ^{6}%
\end{array}%
\right) \allowbreak .
\end{equation}%
Therefore, the matrix $M_{l}M_{l}^{T}$ can be diagonalized by
rotation matrix $R_{l}$ according to: 
\begin{eqnarray}
R_{l}^{T}M_{l}M_{l}^{T}R_{l}&=&\left( 
\begin{array}{ccc}
m_{e}^{2} & 0 & 0 \\ 
0 & m_{\mu }^{2} & 0 \\ 
0 & 0 & m_{\tau }^{2}%
\end{array}%
\right) ,\hspace{1cm}R_{l}=\left( 
\begin{array}{ccc}
1 & 0 & 0 \\ 
0 & \cos \theta _{l} & -\sin \theta _{l} \\ 
0 & \sin \theta _{l} & \cos \theta _{l}%
\end{array}%
\right) ,\hspace{1cm}\tan \theta _{l}\simeq -\frac{z_{1}}{z_{2}}.
\label{Mlred}
\end{eqnarray}%
The charged lepton masses are approximately given by 
\begin{equation}
m_{e}=x_{1}\lambda ^{8}\frac{v}{\sqrt{2}},\hspace{1cm}m_{\mu }\simeq \frac{%
\left\vert y_{1}z_{2}-y_{2}z_{1}\right\vert }{\sqrt{z_{1}^{2}+z_{2}^{2}}}%
\lambda ^{5}\frac{v}{\sqrt{2}},\hspace{1cm}m_{\tau }\simeq \sqrt{%
z_{1}^{2}+z_{2}^{2}}\lambda ^{3}\frac{v}{\sqrt{2}}.  \label{ml}
\end{equation}%
From the neutrino Yukawa terms it follows that the full $5\times 5$ neutrino
mass matrix is 
\begin{equation}
M_{\nu }=\left( 
\begin{array}{cc}
0_{3\times 3} & M_{\nu }^{D} \\ 
\left( M_{\nu }^{D}\right) ^{T} & M_{R}%
\end{array}%
\right) ,  \label{Mnu}
\end{equation}%
where: 
\begin{equation}
M_{\nu }^{D}=\left( 
\begin{array}{cc}
\lambda ^{3}\varepsilon _{11}^{\left( \nu \right) }\frac{v_{2}}{\sqrt{2}} & 
\lambda ^{3}\varepsilon _{12}^{\left( \nu \right) }\frac{v_{2}}{\sqrt{2}} \\ 
\varepsilon _{21}^{\left( \nu \right) }\frac{v_{1}}{\sqrt{2}} & \varepsilon
_{22}^{\left( \nu \right) }\frac{v_{3}}{\sqrt{2}} \\ 
\varepsilon _{31}^{\left( \nu \right) }\frac{v_{1}}{\sqrt{2}} & \varepsilon
_{33}^{\left( \nu \right) }\frac{v_{3}}{\sqrt{2}}%
\end{array}%
\right) =\left( 
\begin{array}{cc}
A & F \\ 
B & E \\ 
C & D%
\end{array}%
\right) ,\hspace{1cm}M_{R}=\left( 
\begin{array}{cc}
M_{1} & \frac{1}{2}M_{12} \\ 
\frac{1}{2}M_{12} & M_{2}%
\end{array}%
\right) .
\end{equation}%
Since $\left( M_{R}\right) _{ii}>>v$, the light neutrino mass matrix is
generated through a type I seesaw mechanism and is given by 
\begin{eqnarray}
M_{L} &=&M_{\nu }^{D}M_{R}^{-1}\left( M_{\nu }^{D}\right) ^{T}=\left( 
\begin{array}{cc}
A & F \\ 
B & E \\ 
C & D%
\end{array}%
\right) \left( 
\begin{array}{cc}
-\frac{4M_{2}}{M_{12}^{2}-4M_{1}M_{2}} & \frac{2M_{12}}{%
M_{12}^{2}-4M_{1}M_{2}} \\ 
\frac{2M_{12}}{M_{12}^{2}-4M_{1}M_{2}} & -\frac{4M_{1}}{%
M_{12}^{2}-4M_{1}M_{2}}%
\end{array}%
\right) \allowbreak \left( 
\begin{array}{ccc}
A & B & C \\ 
F & E & D%
\end{array}%
\right)  \notag \\
&=&\left( 
\begin{array}{ccc}
-\frac{4\left( M_{2}A^{2}-M_{12}AF+M_{1}F^{2}\right) }{M_{12}^{2}-4M_{1}M_{2}%
} & \frac{2\left( BFM_{12}-2ABM_{2}-2FEM_{1}+AEM_{12}\right) }{%
M_{12}^{2}-4M_{1}M_{2}} & \frac{2\left(
CFM_{12}-2ACM_{2}-2FDM_{1}+ADM_{12}\right) }{M_{12}^{2}-4M_{1}M_{2}} \\ 
\frac{2\left( BFM_{12}-2ABM_{2}-2FEM_{1}+AEM_{12}\right) }{%
M_{12}^{2}-4M_{1}M_{2}} & -\frac{4\left(
M_{2}B^{2}-M_{12}BE+M_{1}E^{2}\right) }{M_{12}^{2}-4M_{1}M_{2}} & \frac{%
2\left( BDM_{12}-2BCM_{2}+CEM_{12}-2DEM_{1}\right) }{M_{12}^{2}-4M_{1}M_{2}}
\\ 
\frac{2\left( CFM_{12}-2ACM_{2}-2FDM_{1}+ADM_{12}\right) }{%
M_{12}^{2}-4M_{1}M_{2}} & \frac{2\left(
BDM_{12}-2BCM_{2}+CEM_{12}-2DEM_{1}\right) }{M_{12}^{2}-4M_{1}M_{2}} & -%
\frac{4\left( M_{2}C^{2}-M_{12}CD+M_{1}D^{2}\right) }{M_{12}^{2}-4M_{1}M_{2}}%
\end{array}%
\right) \allowbreak  \notag \\
&=&\left( 
\begin{array}{ccc}
W^{2} & WX\cos \varphi & WY\cos \left( \varphi -\varrho \right) \\ 
WX\cos \varphi & X^{2} & XY\cos \varrho \\ 
WY\cos \left( \varphi -\varrho \right) & XY\cos \varrho & Y^{2}%
\end{array}%
\right) .
\end{eqnarray}%
In order to demonstrate these structures can be fit to the data, we set $%
\varphi =\varrho $ for simplicity, to obtain 
\begin{equation}
M_{L}=\left( 
\begin{array}{ccc}
W^{2} & \kappa WX & WY \\ 
\kappa WX & X^{2} & \kappa XY \\ 
WY & \kappa XY & Y^{2}%
\end{array}%
\right) ,\hspace{2cm}\kappa =\cos \varphi .
\end{equation}

Assuming that the neutrino Yukawa couplings are real, we find that for the
normal (NH) and inverted (IH) mass hierarchies, the light neutrino mass
matrix is diagonalized by a rotation matrix $R_{\nu }$, according to 
\begin{eqnarray}
R_{\nu }^{T}M_{L}R_{\nu } &=&\left( 
\begin{array}{ccc}
0 & 0 & 0 \\ 
0 & m_{\nu _{2}} & 0 \\ 
0 & 0 & m_{\nu _{3}}%
\end{array}%
\right) \allowbreak,\hspace{0.7cm}R_{\nu }=\left( 
\begin{array}{ccc}
-\frac{Y}{\sqrt{W^{2}+Y^{2}}} & \frac{W}{\sqrt{W^{2}+Y^{2}}}\sin \theta
_{\nu } & \frac{W}{\sqrt{W^{2}+Y^{2}}}\cos \theta _{\nu } \\ 
0 & \cos \theta _{\nu } & -\sin \theta _{\nu } \\ 
\frac{W}{\sqrt{W^{2}+Y^{2}}} & \frac{Y}{\sqrt{W^{2}+Y^{2}}}\sin \theta _{\nu
} & \frac{Y}{\sqrt{W^{2}+Y^{2}}}\cos \theta _{\nu }%
\end{array}%
\right) ,\hspace{0.7cm}\mbox{for NH}  \label{NeutrinomassNH} \\
\tan \theta _{\nu } &=&-\sqrt{\frac{m_{3}-X^{2}}{X^{2}-m_{2}}}\allowbreak ,%
\hspace{0.7cm}m_{\nu _{1}}=0,\hspace{0.7cm}m_{\nu _{2,3}}=\frac{%
W^{2}+X^{2}+Y^{2}}{2}\mp \frac{\sqrt{\left( W^{2}-X^{2}+Y^{2}\right)
^{2}-4\kappa ^{2}X^{2}\left( W^{2}+Y^{2}\right) }}{2}.  \notag
\end{eqnarray}%
\begin{eqnarray}
R_{\nu }^{T}M_{L}R_{\nu } &=&\left( 
\begin{array}{ccc}
m_{\nu _{1}} & 0 & 0 \\ 
0 & m_{\nu _{2}} & 0 \\ 
0 & 0 & 0%
\end{array}%
\right) \allowbreak ,\hspace{0.7cm}R_{\nu }=\left( 
\begin{array}{ccc}
\frac{W}{\sqrt{W^{2}+Y^{2}}} & -\frac{Y}{\sqrt{W^{2}+Y^{2}}}\sin \theta
_{\nu } & -\frac{Y}{\sqrt{W^{2}+Y^{2}}}\cos \theta _{\nu } \\ 
0 & \cos \theta _{\nu } & -\sin \theta _{\nu } \\ 
\frac{Y}{\sqrt{W^{2}+Y^{2}}} & \frac{W}{\sqrt{W^{2}+Y^{2}}}\sin \theta _{\nu
} & \frac{W}{\sqrt{W^{2}+Y^{2}}}\cos \theta _{\nu }%
\end{array}%
\right) \allowbreak ,\hspace{0.7cm}\mbox{for IH}  \label{NeutrinomassIH} \\
\allowbreak \tan \theta _{\nu } &=&-\sqrt{\frac{m_{2}-X^{2}}{X^{2}-m_{1}}}%
\allowbreak ,\hspace{0.7cm}m_{\nu _{1,2}}=\frac{W^{2}+X^{2}+Y^{2}}{2}\mp 
\frac{1}{2}\sqrt{\left( W^{2}-X^{2}+Y^{2}\right) ^{2}-4\kappa
^{2}X^{2}\left( W^{2}+Y^{2}\right) },\hspace{0.7cm}m_{\nu _{3}}=0.  \notag
\end{eqnarray}%
The smallness of the active neutrinos masses is a consequence of their
scaling with the inverse of the large Majorana neutrino masses, as expected
from the type I seesaw mechanism implemented in our model.

With the rotation matrices in the charged lepton sector $R_{l}$, Eq. (\ref{Mlred}), and the neutrino sector $R_{\nu }$, Eqs. (\ref{NeutrinomassNH})
and (\ref{NeutrinomassIH}) for NH and IH, respectively, we obtain the PMNS
mixing matrix 
\begin{equation}
U=R_{l}^{T}R_{\nu }=\left\{ 
\begin{array}{l}
\left( 
\begin{array}{ccc}
-\frac{Y}{\sqrt{W^{2}+Y^{2}}} & \frac{W}{\sqrt{W^{2}+Y^{2}}}\sin \theta
_{\nu } & \frac{W}{\sqrt{W^{2}+Y^{2}}}\cos \theta _{\nu } \\ 
&  &  \\ 
\frac{W}{\sqrt{W^{2}+Y^{2}}}\sin \theta _{l} & \cos \theta _{l}\cos \theta
_{\nu }+\frac{Y}{\sqrt{W^{2}+Y^{2}}}\sin \theta _{l}\sin \theta _{\nu } & 
\frac{Y}{\sqrt{W^{2}+Y^{2}}}\cos \theta _{\nu }\sin \theta _{l}-\cos \theta
_{l}\sin \theta _{\nu } \\ 
&  &  \\ 
\frac{W}{\sqrt{W^{2}+Y^{2}}}\cos \theta _{l} & \frac{Y}{\sqrt{W^{2}+Y^{2}}}%
\cos \theta _{l}\sin \theta _{\nu }-\cos \theta _{\nu }\sin \theta _{l} & 
\sin \theta _{l}\sin \theta _{\nu }+\frac{Y}{\sqrt{W^{2}+Y^{2}}}\cos \theta
_{l}\cos \theta _{\nu }%
\end{array}%
\right) \allowbreak \ \ \ \mbox{for NH}, \\ 
\\ 
\left( 
\begin{array}{ccc}
\frac{W}{\sqrt{W^{2}+Y^{2}}} & -\frac{Y}{\sqrt{W^{2}+Y^{2}}}\sin \theta
_{\nu } & -\frac{Y}{\sqrt{W^{2}+Y^{2}}}\cos \theta _{\nu } \\ 
&  &  \\ 
\frac{Y}{\sqrt{W^{2}+Y^{2}}}\sin \theta _{l} & \cos \theta _{l}\cos \theta
_{\nu }+\frac{W}{\sqrt{W^{2}+Y^{2}}}\sin \theta _{\nu }\sin \theta _{l} & 
\frac{W}{\sqrt{X^{2}+Y^{2}}}\sin \theta _{l}\cos \theta _{\nu }-\cos \theta
_{l}\sin \theta _{\nu } \\ 
&  &  \\ 
\frac{Y}{\sqrt{W^{2}+Y^{2}}}\cos \theta _{l} & \frac{W}{\sqrt{W^{2}+Y^{2}}}%
\sin \theta _{\nu }\cos \theta _{l}-\cos \theta _{\nu }\sin \theta _{l} & 
\sin \theta _{l}\sin \theta _{\nu }+\frac{W}{\sqrt{W^{2}+Y^{2}}}\cos \theta
_{l}\cos \theta _{\nu }%
\end{array}%
\right) \allowbreak \ \ \ \mbox{for IH}.%
\end{array}%
\right.  \label{UPMNS}
\end{equation}
By comparing with the standard parametrization we derive the mixing angles
for NH and IH 
\begin{eqnarray}
\sin ^{2}\theta _{12} &=&\frac{W^{2}\sin ^{2}\theta _{\nu }}{Y^{2}+\left(
1-\cos ^{2}\theta _{\nu }\right) W^{2}},\hspace{1cm}\hspace{1cm}\sin
^{2}\theta _{13}=\frac{W^{2}\cos ^{2}\theta _{\nu }}{W^{2}+Y^{2}},  \notag \\
\sin ^{2}\theta _{23} &=&\frac{\left( \sqrt{W^{2}+Y^{2}}\sin \theta _{\nu
}\cos \theta _{l}-Y\cos \theta _{\nu }\sin \theta _{l}\right) ^{2}}{\left(
1-\cos ^{2}\theta _{\nu }\right) W^{2}+Y^{2}}\ ,\ \ \ \ \ \ \ \ \mbox{for NH}
\label{mixinganglesNH}
\end{eqnarray}

\begin{eqnarray}
\sin ^{2}\theta _{12} &=&\frac{Y^{2}\sin ^{2}\theta _{\nu }}{W^{2}+\left(
1-\cos ^{2}\theta _{\nu }\right) Y^{2}},\hspace{1cm}\hspace{1cm}\sin
^{2}\theta _{13}=\frac{Y^{2}\cos ^{2}\theta _{\nu }}{W^{2}+Y^{2}},  \notag \\
\sin ^{2}\theta _{23} &=&\frac{\left( \sqrt{W^{2}+Y^{2}}\sin \theta _{\nu
}\cos \theta _{l}-W\cos \theta _{\nu }\sin \theta _{l}\right) ^{2}}{\left(
1-\cos ^{2}\theta _{\nu }\right) Y^{2}+W^{2}}\ ,\ \ \ \ \ \ \ \ 
\mbox{for
IH.}  \label{mixingnaglesIH}
\end{eqnarray}

We further simplify the analysis by considering 
\begin{equation}
x_{1}=y_{2}=z_{1},
\end{equation}%
so that the charged lepton masses will be determined by three dimensionless
effective parameters, i.e, $x_{1}$, $y_{1}$ and $z_{2}$, whereas the
neutrino mass squared splittings and neutrino mixing parameters will be
controlled by four dimensionless effective parameters, i.e, $\kappa $, $W$, $X$ and $Y$. Varying the parameters $x_{1}$, $y_{1}$, $z_{2}$, $\kappa$, $W$, $X$ and $Y$, we fit
the charged lepton masses, the neutrino mass squared splittings $\Delta m_{21}^{2}$, $\Delta m_{31}^{2}$ (defined as $\Delta m_{ij}^{2}=m_{i}^{2}-m_{j}^{2}$) and the leptonic mixing angles $\sin^{2}\theta _{12}$, $\sin ^{2}\theta _{13}$ and $\sin ^{2}\theta _{23}$\ to
their experimental values for NH and IH. Therefore the lepton sector of our model
contains seven effective free parameters, i.e., $x_{1}$, $y_{1}$, $z_{2}$, $\kappa $, $W$, $X$ and $Y$, and describes the lepton masses and mixing pattern,
characterized by eight physical observables, i.e., the three charged
lepton masses, the two neutrino mass squared splittings and the three
leptonic mixing angles. The results shown in Table \ref{observablesleptonsector} correspond to the following best-fit values: 
\begin{eqnarray}
\kappa &\simeq &0.45,\hspace{1cm}W\simeq 0.13\,eV^{\frac{1}{2}},\hspace{1cm}%
X\simeq 0.11\,eV^{\frac{1}{2}},\hspace{1cm}Y\simeq 0.18\,eV^{\frac{1}{2}}, 
\notag \\
x_{1} &\simeq &0.42,\hspace{1cm}y_{1}\simeq 1.39,\hspace{1cm}z_{2}\simeq
0.77,\ \ \ \ \ \ \ \ \mbox{for NH,}  \label{fitresultsNH}
\end{eqnarray}%
\begin{eqnarray}
\kappa &\simeq &4.03\times 10^{-3},\hspace{1cm}W\simeq 0.18\,eV^{\frac{1}{2}%
},\hspace{1cm}X\simeq 0.22\,eV^{\frac{1}{2}},\hspace{1cm}Y\simeq 0.13\,eV^{%
\frac{1}{2}},  \notag \\
x_{1} &\simeq &0.42,\hspace{1cm}y_{1}\simeq 1.38,\hspace{1cm}z_{2}\simeq
0.78,\ \ \ \ \ \ \ \ \mbox{for IH.}  \label{fitresultsIH}
\end{eqnarray}
Using the best-fit values given above, we obtain the following neutrino
masses for NH and IH 
\begin{equation}
m_{1}=0,\hspace{1cm}m_{2}\approx 9\mbox{meV},\hspace{1cm}m_{3}\approx 50%
\mbox{meV},\ \ \ \ \ \ \ \ \mbox{for NH,}  \label{mnuNH}
\end{equation}%
\begin{equation}
m_{1}\approx 49\mbox{meV},\hspace{1cm}m_{2}\approx 50\mbox{meV},\hspace{1cm}%
m_{3}=0,\ \ \ \ \ \ \ \ \mbox{for IH.}  \label{mnuIH}
\end{equation}%
The obtained and experimental values of the observables in the lepton sector
are shown in Table \ref{observablesleptonsector}. Given that the lightest neutrino is predicted to be massless in our model, the neutrino masses are hierarchical, which puts the overall neutrino mass scale below the current experimental reach (the same applies to the cosmological bound $\sum^3_{k=1}m_{\nu_k}<0.23$ eV on the sum of the neutrino masses \cite{Ade:2013zuv,Bilenky:2014ema}). Therefore, our model fulfills the cosmological contraints on neutrino masses for both normal and inverted hierarchies. 
 
The experimental values of
the charged lepton masses, which are given at the $M_{Z}$ scale, have been
taken from Ref. \cite{Bora:2012tx} , whereas the experimental values of the
neutrino mass squared splittings and leptonic mixing angles for both NH and
IH, are taken from Ref. \cite{Forero:2014bxa}. The obtained charged lepton
masses, neutrino mass squared splittings and lepton mixing angles are in
excellent agreement with the experimental data, showing that the model can
perfectly account for all the observables in the lepton sector. We recall
that for the sake of simplicity, we assumed all leptonic parameters to be
real and further restricted the set of parameters, but a non-vanishing CP
violating phase in the PMNS mixing matrix can be generated by allowing one
or several parameters in the neutrino mass matrix of Eq. (\ref{Mnu}) to be
complex.

\begin{table}[tbh]
\begin{center}
\begin{tabular}{c|l|l}
\hline\hline
Observable & Model value & Experimental value \\ \hline
$m_{e}(MeV)$ & \quad $0.487$ & \quad $0.487$ \\ \hline
$m_{\mu }(MeV)$ & \quad $102.8$ & \quad $102.8\pm 0.0003$ \\ \hline
$m_{\tau }(GeV)$ & \quad $1.75$ & \quad $1.75\pm 0.0003$ \\ \hline
$\Delta m_{21}^{2}$($10^{-5}$eV$^{2}$) (NH) & \quad $7.60$ & \quad $%
7.60_{-0.18}^{+0.19}$ \\ \hline
$\Delta m_{31}^{2}$($10^{-3}$eV$^{2}$) (NH) & \quad $2.48$ & \quad $%
2.48_{-0.07}^{+0.05}$ \\ \hline
$\sin ^{2}\theta _{12}$ (NH) & \quad $0.323$ & \quad $0.323\pm 0.016$ \\ 
\hline
$\sin ^{2}\theta _{23}$ (NH) & \quad $0.567$ & \quad $%
0.567_{-0.128}^{+0.032} $ \\ \hline
$\sin ^{2}\theta _{13}$ (NH) & \quad $0.0234$ & \quad $0.0234\pm 0.0020$ \\ 
\hline
$\Delta m_{21}^{2}$($10^{-5}$eV$^{2}$) (IH) & \quad $7.60$ & \quad $%
7.60_{-0.18}^{+0.19}$ \\ \hline
$\Delta m_{13}^{2}$($10^{-3}$eV$^{2}$) (IH) & \quad $2.48$ & \quad $%
2.48_{-0.06}^{+0.05}$ \\ \hline
$\sin ^{2}\theta _{12}$ (IH) & \quad $0.323$ & \quad $0.323\pm 0.016$ \\ 
\hline
$\sin ^{2}\theta _{23}$ (IH) & \quad $0.573$ & \quad $%
0.573_{-0.043}^{+0.025} $ \\ \hline
$\sin ^{2}\theta _{13}$ (IH) & \quad $0.0240$ & \quad $0.0240\pm 0.0019$ \\ 
\hline
\end{tabular}%
\end{center}
\caption{Model and experimental values of the lepton sector observables, for
normal (NH) and inverted (IH) hierarchies. }
\label{observablesleptonsector}
\end{table}
We can now predict the amplitude for neutrinoless double beta (%
$0\nu\beta\beta$) decay in our model, which is proportional to the effective Majorana
neutrino mass 
\begin{equation}
m_{\beta\beta}=\biggl|\sum_kU^2_{ek}m_{\nu_k}\biggr|,  \label{mee}
\end{equation}
where $U^2_{ek}$ and $m_{\nu_k}$ are the PMNS mixing matrix elements and the
Majorana neutrino masses, respectively.

Then, from Eqs. (\ref{UPMNS}) and (\ref{fitresultsNH})-(\ref{mnuIH}), we
predict the following effective neutrino mass for both hierarchies: 
\begin{equation}
m_{\beta \beta }=\left\{ 
\begin{array}{l}
4\ \mbox{meV}\ \ \ \ \ \ \ \mbox{for \ \ \ \ NH} \\ 
50\ \mbox{meV}\ \ \ \ \ \ \ \mbox{for \ \ \ \ IH} \\ 
\end{array}%
\right.   \label{eff-mass-pred}
\end{equation}%
This is beyond the reach of the present and forthcoming $0\nu\beta\beta$ decay experiments. The present best upper limit on this parameter $%
m_{\beta \beta }\leq 160$ meV comes from the recently quoted EXO-200
experiment \cite{Auger:2012ar,Auty:2013yya} $T_{1/2}^{0\nu \beta \beta }(^{136}\mathrm{Xe}%
)\geq 1.6\times 10^{25}$ yr at the 90 \% CL. This limit will be improved
within the not too distant future. The GERDA experiment \cite%
{Abt:2004yk,Ackermann:2012xja} is currently moving to \textquotedblleft
phase-II\textquotedblright , at the end of which it is expected to reach %
\mbox{$T^{0\nu\beta\beta}_{1/2}(^{76}{\rm Ge}) \geq 2\times 10^{26}$ yr},
corresponding to $m_{\beta \beta }\leq 100$ MeV. A bolometric CUORE
experiment, using ${}^{130}\text{Te}$ \cite{Alessandria:2011rc}, is currently under
construction. Its estimated sensitivity is around $T_{1/2}^{0\nu \beta \beta
}(^{130}\mathrm{Te})\sim 10^{26}$ yr corresponding to \mbox{$m_{\beta\beta}%
\leq 50$ meV.} There are also proposals for ton-scale next-to-next
generation $0\nu \beta \beta $ experiments with $^{136}$Xe \cite%
{KamLANDZen:2012aa,Albert:2014fya} and $^{76}$Ge \cite%
{Abt:2004yk,Guiseppe:2011me} claiming sensitivities over $T_{1/2}^{0\nu
\beta \beta }\sim 10^{27}$ yr, corresponding to $m_{\beta \beta }\sim 12-30$
meV. For recent experimental reviews, see for example Ref. \cite%
{Bilenky:2014uka} and references therein. Thus, according to Eq. (\ref%
{eff-mass-pred}) our model predicts $T_{1/2}^{0\nu \beta \beta }$ at the
level of sensitivities of the next generation or next-to-next generation $0\nu \beta \beta $ experiments. 


\section{Scalar phenomenology \label{scalarpheno}}

The renormalizable scalar potential involving only the $SU(2)$ doublets $%
\phi _{i} $ is 
\begin{align*}
V(\phi _{i})& =-\sum_{i=1}^{2}\mu _{i}^{2}(\phi _{i}^{\dagger }\phi
_{i})+\sum_{i=1}^{2}\kappa _{i}(\phi _{i}^{\dagger }\phi _{i})^{2}, \\
V(\phi _{1},\phi _{2})& =\gamma _{12}(\phi _{1}^{\dagger }\phi _{1})(\phi
_{2}^{\dagger }\phi _{2})+\kappa _{12}(\phi _{1}^{\dagger }\phi _{2})(\phi
_{2}^{\dagger }\phi _{1}), \\
V(\xi ,\chi ,\zeta ,\phi _{i})& =\left( \lambda _{\xi }(\xi \xi
)_{1}+\lambda _{\chi }(\chi ^{\dagger }\chi )+\lambda _{\zeta }(\zeta
^{\dagger }\zeta )\right) \sum_{i=1}^{2}\lambda _{1i}(\phi _{i}^{\dagger
}\phi _{i}),
\end{align*}%
whereas the remaining terms are 
\begin{align*}
V(\xi )& =-\mu _{\xi }^{2}(\xi \xi )_{1}+\gamma _{\xi ,3}(\xi \xi )_{2}\xi
+\kappa _{\xi ,1}(\xi \xi )_{1}(\xi \xi )_{1}+\kappa _{\xi ,2}(\xi \xi
)_{2}(\xi \xi )_{2}, \\
V(\chi )& =-\mu _{\chi }^{2}(\chi ^{\dagger }\chi )+\kappa _{\chi }(\chi
^{\dagger }\chi )^{2}, \\
V(\zeta )& =-\mu _{\zeta }^{2}(\zeta ^{\dagger }\zeta )+\kappa _{\zeta
}(\zeta ^{\dagger }\zeta )^{2}, \\
V(\xi ,\chi ,\zeta )& =\lambda _{2}(\xi \xi )_{1}(\chi ^{\dagger }\chi
)+\lambda _{3}(\xi \xi )_{1}(\zeta ^{\dagger }\zeta )+\lambda _{4}(\zeta
^{\dagger }\zeta )(\chi ^{\dagger }\chi ).
\end{align*}%
To obtain a viable low-energy model with one CP-odd and one charged
Goldstone boson, we consider the following soft breaking terms: 
\begin{align}
V_{\text{soft}}(\zeta ,\chi )& =-\mu _{\chi \zeta }^{2}(\zeta \chi +\zeta
^{\dagger }\chi ^{\dagger }), \\
V_{\text{soft}}(\phi _{i},\phi _{j})& =-\mu _{12}^{2}\left[ \left( \phi
_{1}^{\dagger }\phi _{2}\right) +\left( \phi _{2}^{\dagger }\phi _{1}\right) %
\right] .
\end{align}%
The mass matrices of the low-energy CP-even neutral scalars $\rho _{1,2}$ ,
CP-odd neutral scalars $\eta _{1,2}$ and charged scalars $\varphi
_{1,2}^{\pm }$ can be written as 
\begin{eqnarray}
M_{1} &=&\frac{1}{2}\left( 
\begin{array}{cc}
2\kappa _{1}v_{1}^{2}+\frac{v_{2}}{v_{1}}\mu _{12}^{2} & \gamma
v_{1}v_{2}-\mu _{12}^{2} \\ 
\gamma v_{1}v_{2}-\mu _{12}^{2} & 2\kappa _{2}v_{2}^{2}+\frac{v_{1}}{v_{2}}%
\mu _{12}^{2}%
\end{array}%
\right) ,  \notag \\
M_{2} &=&\frac{\mu _{12}^{2}}{2}\left( 
\begin{array}{cc}
\frac{v_{2}}{v_{1}} & -1 \\ 
-1 & \frac{v_{1}}{v_{2}}%
\end{array}%
\right) ,  \notag \\
M_{3} &=&\frac{\mu _{12}^{2}+\kappa _{12}v_{1}v_{2}}{2}\left( 
\begin{array}{cc}
\frac{v_{2}}{v_{1}} & -1 \\ 
-1 & \frac{v_{1}}{v_{2}}%
\end{array}%
\right).
\end{eqnarray}

The physical low-energy scalar mass eigenstates are connected with the weak
scalar states by the following relations \cite%
{Branco:2011iw,Bhattacharyya:2015nca} 
\begin{eqnarray}  \label{higgstrafo}
\left( 
\begin{array}{c}
h \\ 
H%
\end{array}%
\right) &=&\left( 
\begin{array}{cc}
\sin \alpha & -\cos \alpha \\ 
-\cos \alpha & -\sin \alpha%
\end{array}%
\right) \left( 
\begin{array}{c}
\rho _{1} \\ 
\rho _{2}%
\end{array}%
\right) ,\hspace{1cm}\hspace{1cm}\tan 2\alpha =\frac{2\left( \gamma
v_{1}v_{2}-\mu _{12}^{2}\right) }{2\left( \kappa _{1}v_{1}^{2}-\kappa
_{2}v_{2}^{2}\right) +\mu _{12}^{2}\left( \frac{v_{2}}{v_{1}}-\frac{v_{1}}{%
v_{2}}\right) }, \\
\left( 
\begin{array}{c}
\pi ^{0} \\ 
A^{0}%
\end{array}%
\right) &=&\left( 
\begin{array}{cc}
\cos \beta & \sin \beta \\ 
\sin \beta & -\cos \beta%
\end{array}%
\right) \left( 
\begin{array}{c}
\eta _{1} \\ 
\eta _{2}%
\end{array}%
\right) ,\hspace{1cm}\left( 
\begin{array}{c}
\pi ^{\pm } \\ 
H^{\pm }%
\end{array}%
\right) =\left( 
\begin{array}{cc}
\cos \beta & \sin \beta \\ 
\sin \beta & -\cos \beta%
\end{array}%
\right) \left( 
\begin{array}{c}
\varphi _{1}^{\pm } \\ 
\varphi _{2}^{\pm }%
\end{array}%
\right) ,\hspace{1cm}\tan \beta =\frac{v_{2}}{v_{1}}  \notag
\end{eqnarray}%
with the low-energy physical scalar masses given by 
\begin{align}
m_{h}^{2}& =\frac{1}{2v_{1}}\left( \kappa _{1}v_{1}^{3}+\kappa
_{2}v_{1}v_{2}^{2}+\mu _{12}^{2}v_{2}-v_{1}\sqrt{\gamma
^{2}v_{1}^{2}v_{2}^{2}-2\gamma \mu _{12}^{2}v_{1}v_{2}+\kappa
_{1}^{2}v_{1}^{4}-2\kappa _{1}\kappa _{2}v_{1}^{2}v_{2}^{2}+\kappa
_{2}^{2}v_{2}^{4}+\mu _{12}^{4}}\right) , \\
m_{H}^{2}& =\frac{1}{2v_{1}}\left( \kappa _{1}v_{1}^{3}+\kappa
_{2}v_{1}v_{2}^{2}+\mu _{12}^{2}v_{2}+v_{1}\sqrt{\gamma
^{2}v_{1}^{2}v_{2}^{2}-2\gamma \mu _{12}^{2}v_{1}v_{2}+\kappa
_{1}^{2}v_{1}^{4}-2\kappa _{1}\kappa _{2}v_{1}^{2}v_{2}^{2}+\kappa
_{2}^{2}v_{2}^{4}+\mu _{12}^{4}}\right) , \\
m_{A^{0}}^{2}& =\frac{\mu _{12}^{2}}{2}\left( \frac{v_{2}}{v_{1}}+\frac{v_{1}%
}{v_{2}}\right) ,\hspace{1cm}\hspace{1cm}m_{H^{\pm }}^{2}=\frac{\mu
_{12}^{2}+\kappa _{12}v_{1}v_{2}}{2}\left( \frac{v_{2}}{v_{1}}+\frac{v_{1}}{%
v_{2}}\right) .
\end{align}

The physical low-energy scalar spectrum of our model includes two massive
charged Higgses ($H^{\pm }$), one CP-odd Higgs ($A^{0}$) and two neutral
CP-even Higgs ($h,H^{0}$) bosons. The scalar $h$ is identified as the
SM-like $126$ GeV Higgs boson found at the LHC. It it noteworthy that the
neutral $\pi ^{0}$ and charged $\pi ^{\pm }$ Goldstone bosons are associated
with the longitudinal components of the $Z$ and $W^{\pm }$ gauge bosons,
respectively.

\begin{figure}[tbp]
\centering
\subfigure[]{\includegraphics[width=0.42\textwidth]{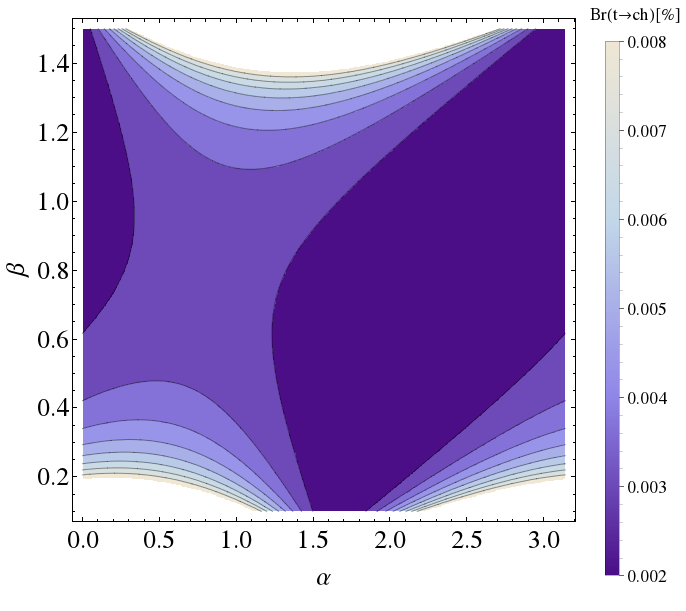}} %
\subfigure[]{\includegraphics[width=0.57\textwidth]{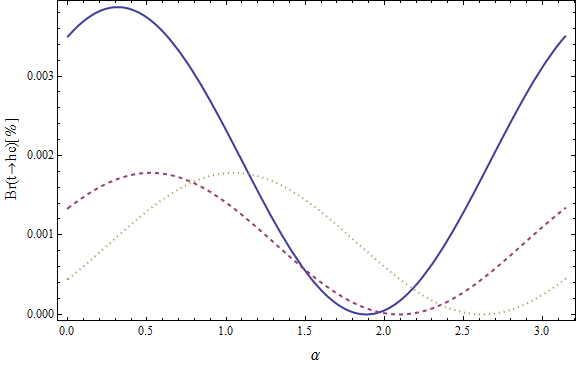}}
\caption{(a)$\text{Br}(t\rightarrow h c)\left[\%\right]$ in the 
$\protect\alpha - \protect\beta$ plane. (b)$\text{Br}(t\rightarrow h c)\left[%
\%\right] $ as a function of $\protect\alpha$ for $\protect\beta=\protect\pi%
/10$ (blue, solid), $\protect\beta=\protect\pi/6$ (red, dashed) and $\protect%
\beta=\protect\pi/3$ (yellow, dotted). The flavor violating $y_{ct}^{h,H}$
couplings are enhanced for small $\protect\beta$ values leading to a
potentially large $\text{Br}(t\rightarrow hc)$ observable at future
experiments.}
\label{fig:tch}
\end{figure}

Thanks to the specific shape of the Yukawa couplings dictated by the
discrete symmetries, the present model is flavor conserving in the down-type and charged lepton sectors because for those sectors we have a special case of Yukawa alignment \cite{Pich:2009sp,Serodio:2011hg, Varzielas:2011jr}. $\phi _{2}$ generates the masses of the first two down-type quark generations,
whereas $\phi _{1}$ is responsible only for the bottom Yukawa, conversely, $\phi _{2}$ is associated only with the electron Yukawa, while $\phi _{1}$ generates the masses of the remaining charged leptons. The Yukawa couplings of both doublets are therefore aligned in these sectors. Due to the lack of
Flavor Changing Neutral Currents (FCNCs) in the down-type sector, tightly
constrained Kaon and B-meson mixings are protected against neutral scalar
contributions. Mixing occurs exclusively in the up-type sector, where both $%
\phi _{1}$ and $\phi _{2}$ couple to the third generation of up-type quarks.
Consequently, top quark FCNCs arise that can be exploited as a probe of new
physics since associated processes are strongly suppressed in the SM.
Explicitly, we obtain the following structures for the up and down-type
Yukawas in the scalar and fermion mass bases using the rotation matrices (%
\ref{quarktrafo}), (\ref{Mlred}), (\ref{higgstrafo}) and the corresponding
transformations of the right handed fields. 
\begin{align}
Y_{h}^{d}& =\left( 
\begin{array}{ccc}
y_{dd}^{h} & y_{ds}^{h} & y_{db}^{h} \\ 
y_{sd}^{h} & y_{ss}^{h} & y_{sb}^{h} \\ 
y_{bd}^{h} & y_{bs}^{h} & y_{bb}^{h}%
\end{array}%
\right) =\sqrt{2}\left( 
\begin{array}{ccc}
-\frac{c_{\alpha }m_{d}}{vs_{\beta }} & 0 & 0 \\ 
0 & -\frac{c_{\alpha }m_{s}}{vs_{\beta }} & 0 \\ 
0 & 0 & \frac{m_{b}s_{\alpha }}{vc_{\beta }}%
\end{array}%
\right) , \\
Y_{H}^{d}& =\left( 
\begin{array}{ccc}
y_{dd}^{H} & y_{ds}^{H} & y_{db}^{H} \\ 
y_{sd}^{H} & y_{ss}^{H} & y_{sb}^{H} \\ 
y_{bd}^{H} & y_{bs}^{H} & y_{bb}^{H}%
\end{array}%
\right) =\sqrt{2}\left( 
\begin{array}{ccc}
-\frac{m_{d}s_{\alpha }}{vs_{\beta }} & 0 & 0 \\ 
0 & -\frac{m_{s}s_{\alpha }}{vs_{\beta }} & 0 \\ 
0 & 0 & -\frac{c_{\alpha }m_{b}}{vc_{\beta }}%
\end{array}%
\right) , \\
Y_{h}^{u}& =\left( 
\begin{array}{ccc}
y_{uu}^{h} & y_{uc}^{h} & y_{ut}^{h} \\ 
y_{cu}^{h} & y_{cc}^{h} & y_{ct}^{h} \\ 
y_{tu}^{h} & y_{tc}^{h} & y_{tt}^{h}%
\end{array}%
\right) \simeq \sqrt{2}\left( 
\begin{array}{ccc}
\frac{m_{u}s_{\alpha }}{vc_{\beta }} & 0 & \frac{m_{t}}{v}V_{tb}V_{ub}\left( 
\frac{c_{\alpha }}{s_{\beta }}+\frac{s_{\alpha }}{c_{\beta }}\right) \\ 
0 & \frac{m_{c}s_{\alpha }}{vc_{\beta }} & \frac{m_{t}}{v}V_{tb}V_{cb}\left( 
\frac{c_{\alpha }}{s_{\beta }}+\frac{s_{\alpha }}{c_{\beta }}\right) \\ 
0 & 0 & \frac{m_{t}}{v}\left( V_{tb}^{2}\frac{s_{\alpha }}{c_{\beta }}-\frac{%
c_{\alpha }}{s_{\beta }}\mathcal{O}(\lambda ^{4})\right)%
\end{array}%
\right) , \\
Y_{H}^{u}& =\left( 
\begin{array}{ccc}
y_{uu}^{H} & y_{uc}^{H} & y_{ut}^{H} \\ 
y_{cu}^{H} & y_{cc}^{H} & y_{ct}^{H} \\ 
y_{tu}^{H} & y_{tc}^{H} & y_{tt}^{H}%
\end{array}%
\right) \simeq \sqrt{2}\left( 
\begin{array}{ccc}
-\frac{c_{\alpha }m_{u}}{vc_{\beta }} & 0 & \frac{m_{t}}{v}%
V_{tb}V_{ub}\left( \frac{s_{\alpha }}{s_{\beta }}-\frac{c_{\alpha }}{%
c_{\beta }}\right) \\ 
0 & -\frac{c_{\alpha }m_{c}}{vc_{\beta }} & \frac{m_{t}}{v}%
V_{tb}V_{cb}\left( \frac{s_{\alpha }}{s_{\beta }}-\frac{c_{\alpha }}{%
c_{\beta }}\right) \\ 
0 & 0 & -\frac{m_{t}}{v}\left( V_{tb}^{2}\frac{c_{\alpha }}{c_{\beta }}+%
\frac{s_{\alpha }}{s_{\beta }}\mathcal{O}(\lambda ^{4})\right)%
\end{array}%
\right),
\end{align}%
with the notations $\sin (x)\equiv s_{x}$, $\cos (x)\equiv c_{x}$ and $\tan
(x)\equiv t_{x}$ and $V_{ij}$ denote the CKM matrix elements. Furthermore, the mixing angles $\alpha $ and $\beta $ are defined in Eq. (%
\ref{higgstrafo}). As in other 2HDMs the couplings depend crucially on the
parameters $\alpha $ and $\beta $, but should comply with the current bounds
if $\tan \beta $ is neither unnaturally large or small, in which cases
deviations from the bottom and top Yukawa couplings with respect to the SM
will become very large. This agrees with our previous statement that the
fermion mass hierarchies and mixing are best explained by $\tan \beta $
values of $\mathcal{O}(1)$. As explained above, FCNCs are absent in the
down-type quark sector since the matrices $Y_{h,H}^{d}$ do not have
off-diagonal entries. The up-type Yukawa couplings $Y_{ut,ct}^{h,H}$,
however, allow for the tree-level decays $t\rightarrow hq$ ($q=u,c$), 
whose branching ratios are currently limited by ATLAS to $\text{%
Br}(t\rightarrow hq)<0.79\%$ @ $95\%$ C.L. \cite{Aad:2014dya} and by CMS to $\text{%
Br}(t\rightarrow cq)<0.56\%$ @ $95\%$ C.L (observed limit) and $\text{Br}(t\rightarrow cq)<0.65^{+0.29}_{-0.19}\%$ (expected limit) \cite{CMS:2014qxa}. Since $y_{ut}$ is negligibly small compared to $y_{ct}$, we consider only the stronger CMS
constraint that can be interpreted as an upper bound on the off-diagonal top
Yukawas to 
\begin{equation}  \label{e:topbound}
\sqrt{|y_{ct}^{h}|^{2}+|y_{ct}^{h}|^{2}}=\frac{\sqrt{2}m_{t}}{v}\sqrt{\left\vert V_{tb}V_{cb}\left( \frac{%
s_{\alpha }}{c_{\beta }}+\frac{c_{\alpha }}{s_{\beta }}\right) \right\vert
^{2}}~<0.14\,,
\end{equation}%
which
translates to 
\begin{equation}  \label{topdecaybound}
\left\vert \frac{c_{\alpha -\beta }}{c_{\beta }s_{\beta }}\right\vert
\lesssim 3.40\,.
\end{equation}%
The $t\rightarrow ch$ channel is particularly interesting since
its branching ratio $\text{Br}(t\rightarrow hc)_{\text{SM}}\simeq 10^{-15}$
\cite{Aad:2014dya} is extremely suppressed in the SM, but can be potentially large in our
model allowing it to be probed at future collider experiments. As shown in
Fig. \ref{fig:tch} our model predictions can reach branching ratios of $%
\mathcal{O}(0.01\%)$ in some regions of the $\alpha -\beta $ plane, allowing
to further constrain our model parameter space with experimental searches
for rare top decays. 

\begin{figure}[tbp]
\centering
\subfigure[]{\includegraphics[width=0.49\textwidth]{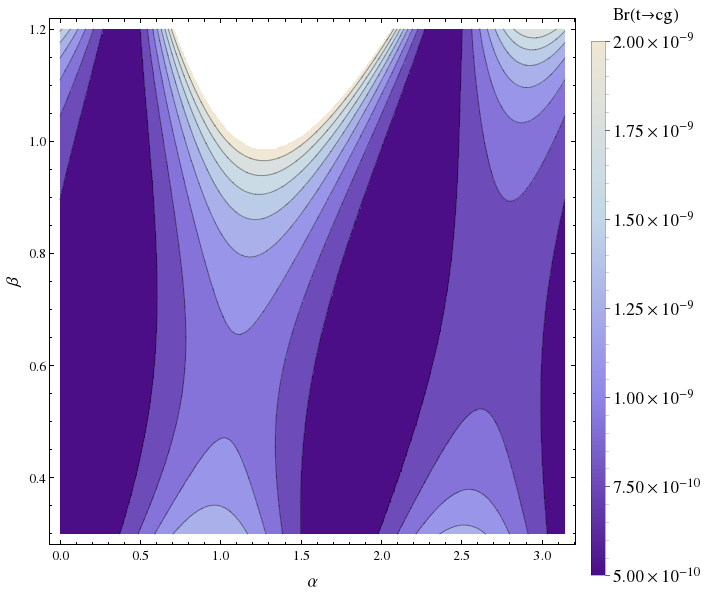}} %
\subfigure[]{\includegraphics[width=0.49\textwidth]{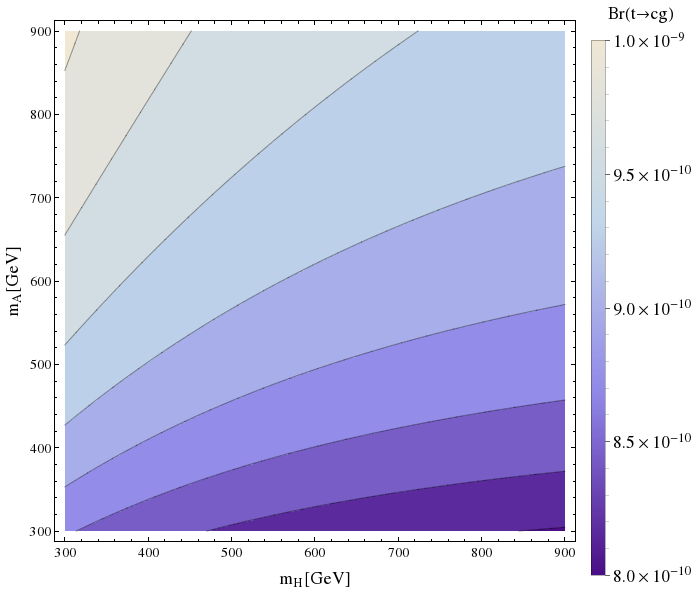}}
\caption{(a)$\text{Br}(t\rightarrow h g)$ in the $\protect%
\alpha - \protect\beta$ plane with $m_H = m_A = 500\,$GeV. (b)$\text{Br}%
(t\rightarrow h g)$ as a function of $m_H$ and $m_A$ for $\protect\alpha=%
\protect\pi/3$ and $\protect\beta=\protect\pi/4$. The decay rate is to a large extent independent of the charged Higgs
mass $m_{H^\pm}$.}
\label{fig:tcg}
\end{figure}

Recently an analysis of up-type FCNCs in the 2HDM type III has been
performed \cite{Kim:2015zla} parametrizing the flavor violating $y^h_{ct}$
coupling as $y_{ct}^h = \frac{1}{v}\lambda_{ct} \sqrt{2 m_t m_c}$ according
to the Cheng--Sher Ansatz \cite{Cheng:1987rs} (this type of FCNC was shown to be remarkably stable under radiative corrections \cite{Cvetic:1998uw}). Focusing on the $cc
\rightarrow tt$ as well as the $t \rightarrow c g$ channels, they find that $%
\lambda_{ct}$ can still take values of up to $10-20$ depending on the
neutral heavy Higgs mass. With $y^h_{ct} \propto \frac{1}{v} V_{cb} V_{tb} 
\sqrt{2} m_t$ our model corresponds to $\lambda_{ct} \approx \frac{1}{2}$
and is therefore well below the critical region. Indeed,
following the analysis of \cite{Atwood:1996vj} we find numerically that the
loop induced decays $t \rightarrow c g$, $t \rightarrow c \gamma$ and $t
\rightarrow c Z$ are several orders of magnitude below the current LHC
sensitivity. Explicitly, varying the free model parameters $\alpha, \beta$
and the scalar masses $m_H, m_A$ and $m_{H^{\pm}}$, we expect the branching
ratios to be approximately 
\begin{align}
\text{Br}(t\rightarrow cg) \sim \mathcal{O}(10^{-9}),\qquad \text{Br}%
(t\rightarrow c \gamma) \sim \mathcal{O}(10^{-12}), \qquad \text{Br}%
(t\rightarrow c Z) \sim \mathcal{O}(10^{-13}),
\end{align}
as opposed to the current upper limits from ATLAS and CMS \cite%
{Chatrchyan:2013nwa,Aad:2015uza} 
\begin{align}
\text{Br}(t\rightarrow cg) < 1.6 \times 10^{-4},\qquad \text{Br}%
(t\rightarrow c \gamma, cZ) <5 \times 10^{-4}.
\end{align}
The largest branching ratio of the three channels, $\text{Br}(t\rightarrow cg)$, is
shown in Fig. \ref{fig:tcg} as a function of $\alpha$ and $\beta$ for fixed $%
m_H$ and $m_A$ (a), as well as for variable $m_H$ and $m_A$ with fixed $%
\alpha$ and $\beta$ (b). As it turns out, the charged Higgs contribution is
tiny and does not affect the prediction for any values of $m_{H^{\pm}}$. 

In the charged lepton sector we obtain 
\begin{align}
Y_h^l &= \sqrt{2}\left( 
\begin{array}{ccc}
y^h_{ee} & y^h_{e\mu} & y^h_{e\tau} \\ 
y^h_{\mu e} & y^h_{\mu\mu} & y^h_{\mu\tau} \\ 
y^h_{\tau e} & y^h_{\tau \mu} & y^h_{\tau \tau}%
\end{array}
\right)= \sqrt{2}\left( 
\begin{array}{ccc}
-\frac{c_{\alpha } m_e}{v s_{\beta }} & 0 & 0 \\ 
0 & \frac{m_{\mu } s_{\alpha }}{v c_{\beta }} & 0 \\ 
0 & 0 & \frac{m_{\tau } s_{\alpha }}{v c_{\beta }}%
\end{array}
\right), \\
Y_H^l &= \sqrt{2}\left( 
\begin{array}{ccc}
y^H_{ee} & y^H_{e\mu} & y^H_{e\tau} \\ 
y^H_{\mu e} & y^H_{\mu\mu} & y^H_{\mu\tau} \\ 
y^H_{\tau e} & y^H_{\tau \mu} & y^H_{\tau \tau}%
\end{array}
\right)= \sqrt{2}\left( 
\begin{array}{ccc}
-\frac{m_e s_{\alpha }}{v s_{\beta }} & 0 & 0 \\ 
0 & -\frac{c_{\alpha } m_{\mu }}{v c_{\beta }} & 0 \\ 
0 & 0 & -\frac{c_{\alpha } m_{\tau }}{v c_{\beta }}%
\end{array}
\right).
\end{align}
The charged leptons are also free of FCNCs due to the lack of off-diagonal
Yukawa couplings. Consequently, the recently reported anomaly in $h\rightarrow \mu\tau$ decays cannot be explained in our present model, even
though it was possible to account for this in other multi-Higgs models with $%
S_3$ or other discrete symmetries \cite{Campos:2014zaa, Sierra:2014nqa,
Heeck:2014qea,Varzielas:2015joa}.

The charged Higgs couplings that are relevant, e.g., for $B^0_{s,d}-%
\overline{B^0}_{s,d}$ mixing and the radiative decays $b \rightarrow q\gamma
~(q=s,d)$, are given by 
\begin{align}
Y_{H^{\pm}}^L&= \sqrt{2}\left( 
\begin{array}{ccc}
y_{du} & y_{dc} & y_{dt} \\ 
y_{su} & y_{sc} & y_{st} \\ 
y_{bu} & y_{bc} & y_{bt}%
\end{array}
\right)= \sqrt{2}\left( 
\begin{array}{ccc}
\frac{V_{ud}}{V_{tb}^2+V_{cb}^2} t_{\beta} \frac{m_u}{v} & -\frac{V_{us}}{%
V_{tb}} t_{\beta} \frac{m_c}{v} & -V_{td}^{\ast} \frac{m_t}{v t_{\beta}} \\ 
\frac{V_{us}}{V_{tb}^2+V_{cb}^2} t_{\beta} \frac{m_u}{v} & \frac{V_{ud}}{%
V_{tb}} t_{\beta} \frac{m_c}{v} & -V_{ts}^{\ast} \frac{m_t}{v t_{\beta} } \\ 
0 & 0 & V_{tb} t_{\beta} \frac{m_t}{v}%
\end{array}
\right), \\
Y_{H^{\pm}}^R&= \sqrt{2}\left( 
\begin{array}{ccc}
y_{ud} & y_{us} & y_{ub} \\ 
y_{cd} & y_{cs} & y_{cb} \\ 
y_{td} & y_{ts} & y_{tb}%
\end{array}
\right)= \sqrt{2}\left( 
\begin{array}{ccc}
V_{ud} \frac{m_d}{v t_{\beta} } & V_{us} \frac{m_s}{v t_{\beta}} & V_{ub}
t_{\beta} \frac{m_b}{v} \\ 
V_{cd} \frac{m_d}{v t_{\beta}} & V_{cs} \frac{m_s}{v t_{\beta}} & V_{cb}
t_{\beta} \frac{m_b}{v} \\ 
V_{td} \frac{m_d}{v t_{\beta}} & V_{ts} \frac{m_s}{v t_{\beta}} & V_{tb}
t_{\beta} \frac{m_b}{v}%
\end{array}
\right),
\end{align}%
\begin{align}
Y_{H^{\pm}}^{e \nu}&=\sqrt{2}\frac{m_e}{v t_{\beta}},\qquad Y_{H^{\pm}}^{\mu
\nu}=\sqrt{2}\frac{m_{\mu}}{v} t_{\beta} \left(c_{\theta_l }-s_{\theta_l
}\right), \qquad Y_{H^{\pm}}^{\tau \nu}=\sqrt{2}\frac{m_{\tau}}{v} t_{\beta}
\left(c_{\theta_l }+s_{\theta_l }\right),
\end{align}
where in the last equation we summed over the neutrino mass eigenstates as
they are usually undetected in typical flavor experiments. Here, the
couplings $y_{bu}$ and $y_{bc}$ that could be used to explain the
outstanding anomaly in $B \rightarrow D^{(\ast)} \tau \nu$ decays \cite%
{Lees:2013uzd} are zero, hence no difference from 2HDMs of type II is to be
expected in these channels.

On the other hand, the charged scalar sector is tightly constrained by $b
\rightarrow s \gamma$ measurements, where the charged scalar $H^{\pm}$ leads
to an additional loop diagram replacing the $W^{\pm}$. Recently a lower
bound of $480\,\text{GeV}$ was placed on the charged Higgs in the 2HDM type
II \cite{Misiak:2015xwa}. Following the analysis of \cite{Trott:2010iz} we
estimate a lower bound on the charged Higgs mass imposed on our model by
constraints on the Wilson coefficients involved in $\text{Br}(b \rightarrow
s \gamma)$. Since $\tan \beta$ drops out in the product of the corresponding
Yukawa couplings $y_{tb} (y_{bt})$ and $y_{ts} (y_{st})$, the prediction is
independent of $\tan \beta$ and the lower limit is roughly $m_{H^{\pm}}
\gtrsim 500\,\text{GeV}$.

\subsection{Constraints from $h\rightarrow \protect\gamma \protect\gamma $ 
\label{gammagamma}}

\begin{figure}[tbp]
\includegraphics[width=1.1\textwidth]{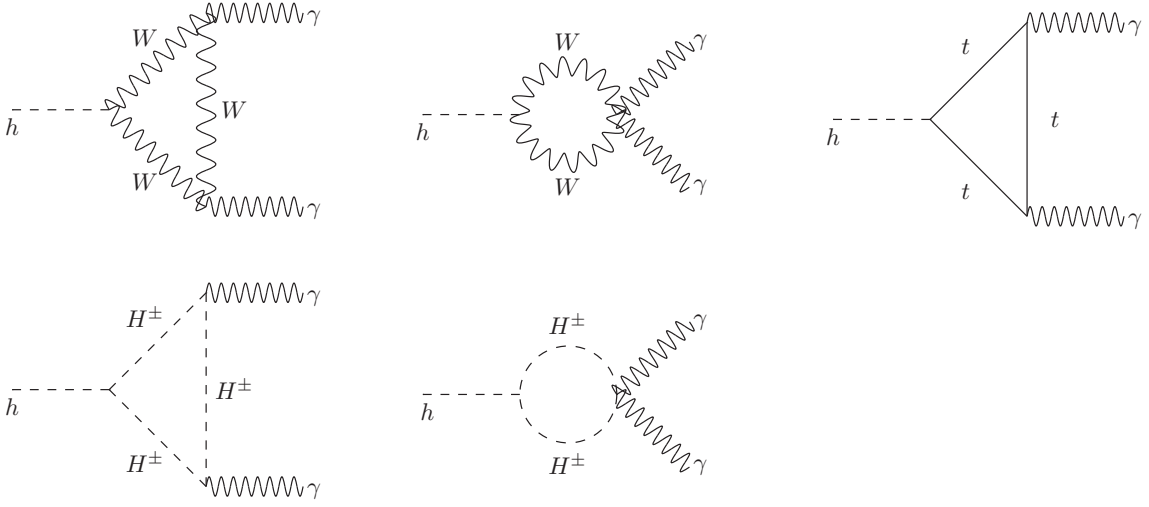}\vspace{-17cm}
\caption{One-loop Feynman diagrams in the Unitary Gauge contributing to the $%
h\rightarrow \protect\gamma \protect\gamma $ decay. }
\label{fig:hgg}
\end{figure}

In our 2HDM the $h\rightarrow \gamma \gamma $ decay receives additional
contributions from loops with charged scalars $H^{\pm }$, as shown in Fig. %
\ref{fig:hgg}, and therefore sets bounds on the masses of these scalars {as
well as on the angles $\alpha$ and $\beta$. }

The explicit form of the $h\rightarrow \gamma \gamma $ decay rate is \cite{Shifman:1979eb,Gavela:1981ri,Kalyniak:1985ct,Gunion:1989we,Spira:1997dg,Djouadi:2005gj,Marciano:2011gm,Wang:2012gm}
\begin{equation}
\Gamma \left( h\rightarrow \gamma \gamma \right) =\frac{\alpha
_{em}^{2}m_{h}^{3}}{256\pi ^{3}v^{2}}\left\vert
\sum_{f}a_{hff}N_{c}Q_{f}^{2}F_{1/2}\left( \varrho _{f}\right)
+a_{hWW}F_{1}\left( \varrho _{W}\right) +\frac{\lambda _{hH^{\pm }H^{\mp }}v%
}{2m_{H^{\pm }}^{2}}F_{0}\left( \varrho _{H^{\pm }}\right) \right\vert
^{2}\,.  \label{e:hggcr}
\end{equation}%
Here $\varrho _{i}$ are the mass ratios $\varrho _{i}=\frac{m_{h}^{2}}{%
4M_{i}^{2}}$, with $M_{i}=m_{f},M_{W}$, and $m_{H^{\pm }}$, $\alpha _{em}$
is the fine structure constant, $N_{C}$ is the color factor ($N_{C}=1$ for
leptons, $N_{C}=3$ for quarks), and $Q_{f}$ is the electric charge of the
fermion in the loop. From the fermion-loop contributions we consider only
the dominant top quark term. Furthermore, $\lambda _{hH^{\pm }H^{\mp }}$ is
the trilinear coupling between the SM-like Higgs and a pair of charged
Higgses, which is given by 
\begin{equation}
\lambda _{hH^{\pm }H^{\mp }}=-\frac{\gamma _{12}+\kappa _{12}}{2}v\sin
2\beta \cos \left( \alpha +\beta \right) .
\end{equation}

Besides that $a_{htt}$ and $a_{hWW}$ are the deviation factors from the SM
Higgs-top quark coupling and the SM Higgs- $W$ gauge boson coupling,
respectively (in the SM these factors are unity). These deviation factors
are given by 
\begin{eqnarray}
a_{htt} &\simeq &\frac{\sin \alpha }{\cos \beta }, \\
a_{hWW} &=&\sin \left( \alpha -\beta \right) ,
\end{eqnarray}%
where in $a_{htt}$ we neglected the contribution suppressed by small CKM
entries.

The dimensionless loop factors $F_{1/2}\left( \varrho \right) $ and $%
F_{1}\left( \varrho \right) $ (for spin-$1/2$ and spin-$1$ particles in the
loop, respectively) are \cite{Gunion:1989we,Djouadi:2005gj} 
\begin{align}
F_{1/2}\left( \varrho \right) & =2\left[ \varrho +\left( \varrho -1\right)
f\left( \varrho \right) \right] \varrho ^{-2}, \\
F_{1}\left( \varrho \right) & =-\left[ 2\varrho ^{2}+3\varrho +3\left(
2\varrho -1\right) f\left( \varrho \right) \right] \varrho ^{-2},  \label{F}
\\
F_{0}\left( \varrho \right) & =-\left[ \varrho -f\left( \varrho \right) %
\right] \varrho ^{-2},
\end{align}%
with 
\begin{equation}
f\left( \varrho \right) =%
\begin{cases}
\arcsin ^{2}\sqrt{\varrho },\hspace{0.5cm}\mathit{for}\hspace{0.2cm}\varrho
\leq 1 \\ 
-\frac{1}{4}\left[ \ln \left( \frac{1+\sqrt{1-\varrho ^{-1}}}{1-\sqrt{%
1-\varrho ^{-1}}}\right) -i\pi \right] ^{2},\hspace{0.5cm}\mathit{for}%
\hspace{0.2cm}\varrho >1.%
\end{cases}%
\end{equation}%
In what follows we determine the constraints that the Higgs diphoton signal
strength imposes on our model. To this end, we introduce the ratio $%
R_{\gamma \gamma }$, which normalizes the $\gamma \gamma $ signal predicted
by our model relative to that of the SM: 
\begin{equation}
R_{\gamma \gamma }=\frac{\sigma \left( pp\rightarrow h\right) \Gamma \left(
h\rightarrow \gamma \gamma \right) }{\sigma \left( pp\rightarrow h\right)
_{SM}\Gamma \left( h\rightarrow \gamma \gamma \right) _{SM}}\simeq
a_{htt}^{2}\frac{\Gamma \left( h\rightarrow \gamma \gamma \right) }{\Gamma
\left( h\rightarrow \gamma \gamma \right) _{SM}}.  \label{R_gamma}
\end{equation}%
The normalization given by Eq. (\ref{R_gamma}) for $h\rightarrow \gamma
\gamma $ was also used in Refs.~\cite%
{Wang:2012gm,Carcamo-Hernandez:2013ypa,Bhattacharyya:2014oka,Fortes:2014dia,Campos:2014zaa,Hernandez:2015xka,Hernandez:2015nga}%
.

The ratio $R_{\gamma \gamma }$ has been measured by CMS and ATLAS with the
best-fit signals \cite{Khachatryan:2014ira,Aad:2014eha} 
\begin{equation*}
R_{\gamma \gamma }^{\text{CMS}}=1.14_{-0.23}^{+0.26}\qquad \text{and}\qquad
R_{\gamma \gamma }^{\text{ATLAS}}=1.17\pm 0.27.
\end{equation*}

\begin{figure}[tbp]
\centering
\subfigure[]{\includegraphics[width=0.59\textwidth]{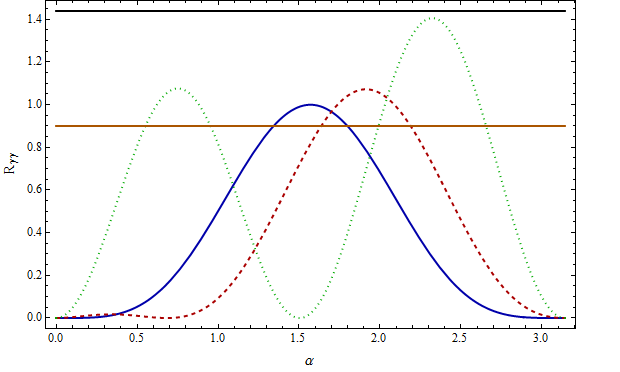}} %
\subfigure[]{\includegraphics[width=0.37%
\textwidth]{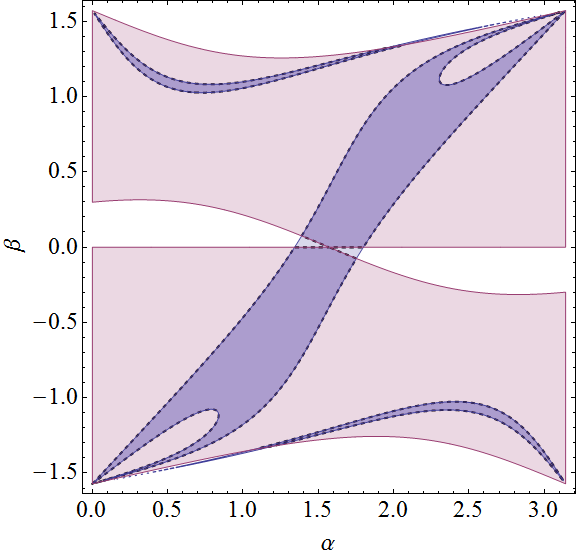}}
\caption{The constraints on the model imposed by keeping $R_{\protect\gamma 
\protect\gamma }$ inside the experimentally allowed 1$\protect\sigma$ range
determined by CMS and ATLAS to be $1.14^{+0.26}_{-0.23}$ and $1.17\pm 0.27 $%
, respectively \protect\cite{Khachatryan:2014ira,Aad:2014eha}. (\protect\ref%
{Rgamma}.(a)) shows the ratio $R_{\protect\gamma \protect\gamma }$ as a
function of the mixing angle $\protect\alpha$ of the CP-even neutral scalars 
$h$ and $H^0$ for $m_{H^{\pm }}=500$ GeV, $\protect\gamma_{12}+\protect\kappa%
_{12}=1$ and different values of the mixing angle $\protect\beta$; the blue,
red and green curves correspond $\protect\beta$ set to $0$, $\frac{\protect%
\pi}{6} $ and $\frac{\protect\pi}{3}$, respectively, and the horizontal
lines are the minimum and maximum values of the ratio $R_{\protect\gamma%
\protect\gamma }$. (\protect\ref{Rgamma}.(b)) shows the allowed region in
the $\protect\alpha$-$\protect\beta$ plane consistent with the Higgs
diphoton decay rate constraint at the LHC, superimposed with the constraint
imposed by Eq.(\protect\ref{topdecaybound}).}
\label{Rgamma}
\end{figure}

Figure (\ref{Rgamma}(a)) shows the sensitivity of the ratio $R_{\gamma
\gamma }$ under variations of the mixing angle $\alpha $ for $m_{H^{\pm
}}=500$ GeV, $\gamma _{12}+\kappa _{12}=1$ and different values of the
mixing angle $\beta $. It follows that as the mixing angle $\beta $ is
increased, the range of $\alpha $ consistent with LHC observations of $%
h\rightarrow \gamma \gamma $ moves away from $\pi/2$. On the other hand, the
decay rate is largely independent of the charged Higgs mass or the sum of
the couplings $\gamma _{12}+\kappa _{12}$, which is consistent with the
contribution mediated by charged scalars to the $h \rightarrow \gamma \gamma$ process being a small
correction. In fact we checked numerically it stays almost constant when $%
m_{H^{\pm }}$ is varied from $500$ GeV to $1$ TeV for fixed values of $%
\alpha ,\beta $, and the quartic couplings of the scalar potential. For the
same values of the charged Higgs mass and quartic couplings, we show in
Figure (\ref{Rgamma}(b)) the Z-shaped allowed region in the $\alpha $-$\beta 
$ plane that is consistent with the Higgs diphoton decay rate constraints at
the LHC, and overlay it with the relatively weak bound in Eq. (\ref%
{topdecaybound}) that arises from top quark FCNCs.

\subsection{$T$ and $S$ parameters \label{TnS}}

The extra scalars affect the oblique corrections of the SM, and these values
are measured in high precision experiments. Consequently, they act as a
further constraint on the validity of our model. The oblique corrections are
parametrized in terms of the two well-known quantities $T$ and $S$. In this
section we calculate one-loop contributions to the oblique parameters $T$
and $S$ defined as \cite{Peskin:1991sw,Altarelli:1990zd,Barbieri:2004qk} 
\begin{equation}
T=\frac{\Pi _{33}\left( q^{2}\right) -\Pi _{11}\left( q^{2}\right) }{\alpha
_{EM}(M_{Z})M_{W}^{2}}\biggl|_{q^{2}=0},\ \ \ \ \ \ \ \ \ \ \ S=\frac{2\sin 2%
{\theta }_{W}}{\alpha _{EM}(M_{Z})}\frac{d\Pi _{30}\left( q^{2}\right) }{%
dq^{2}}\biggl|_{q^{2}=0}.  \label{T-S-definition}
\end{equation}%
$\Pi _{11}\left( 0\right) $, $\Pi _{33}\left( 0\right) $, and $\Pi
_{30}\left( q^{2}\right) $ are the vacuum polarization amplitudes with $%
\{W_{\mu }^{1},W_{\mu }^{1}\}$, $\{W_{\mu }^{3},W_{\mu }^{3}\}$ and $%
\{W_{\mu }^{3},B_{\mu }\}$ external gauge bosons, respectively, where $q$ is
their momentum. We note that in the definitions of the $T$ and $S$
parameters, the new physics is assumed to be heavy when compared to $M_{W}$
and $M_{Z}$.

\begin{figure}[tbh]
\centering
\includegraphics[width=0.7\textwidth]{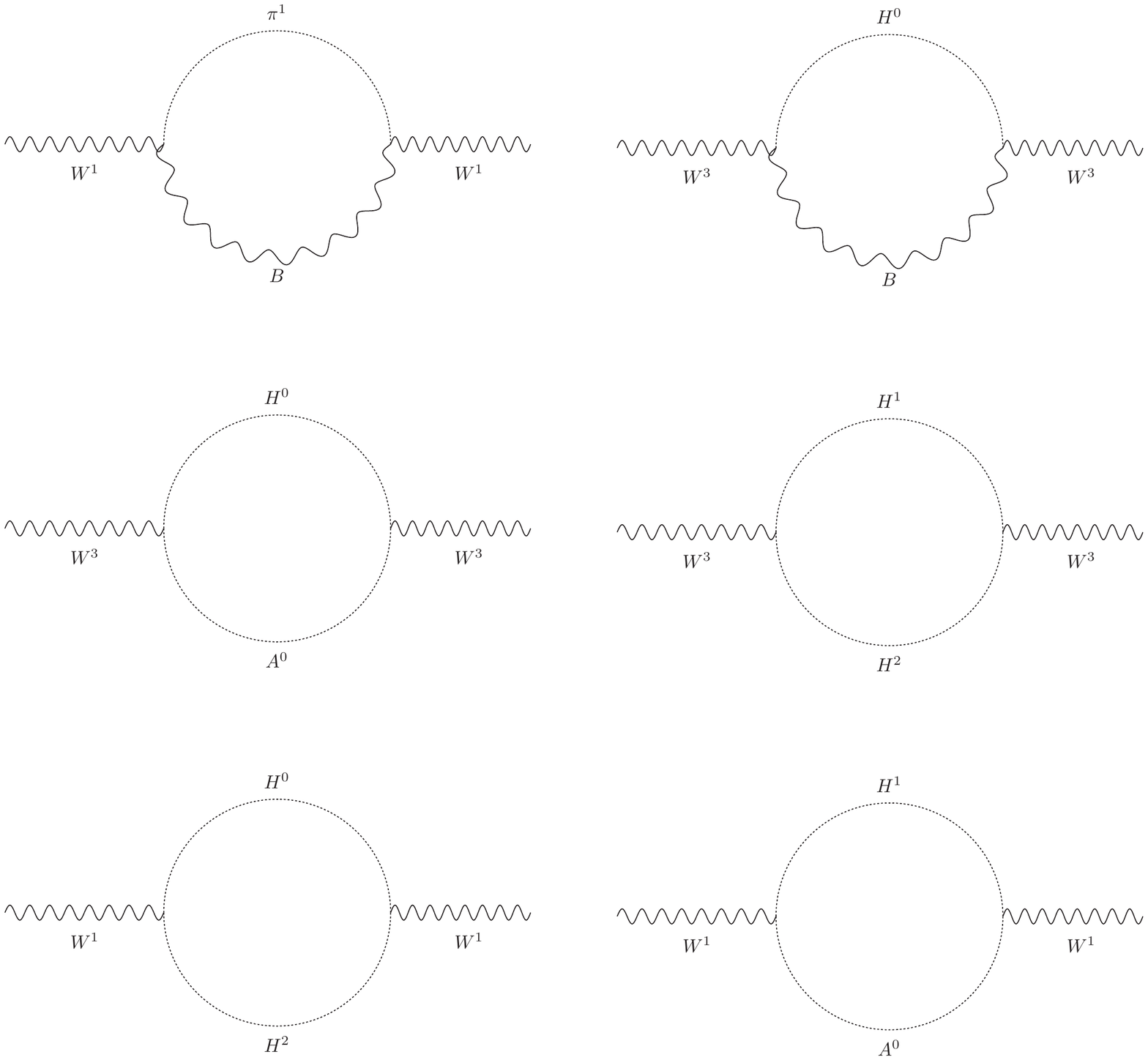}\vspace{-3cm}
\caption{One-loop Feynman diagrams contributing to the $T$ parameter. The
fields $H^1$ and $H^2$ are linear combinations of the charged Higgses $%
H^{\pm}$, similarly to how $W^{\pm}$ gauge bosons are defined in terms of $%
W^1$ and $W^2$. Likewise, the fields $\protect\pi^1$ and $\protect\pi^2$ are
linear combinations of the charged Goldstone bosons $\protect\pi^{\pm}$.}
\label{diagT1}
\end{figure}

\begin{figure}[tbh]
\centering
\vspace{-3cm} \includegraphics[width=0.7\textwidth]{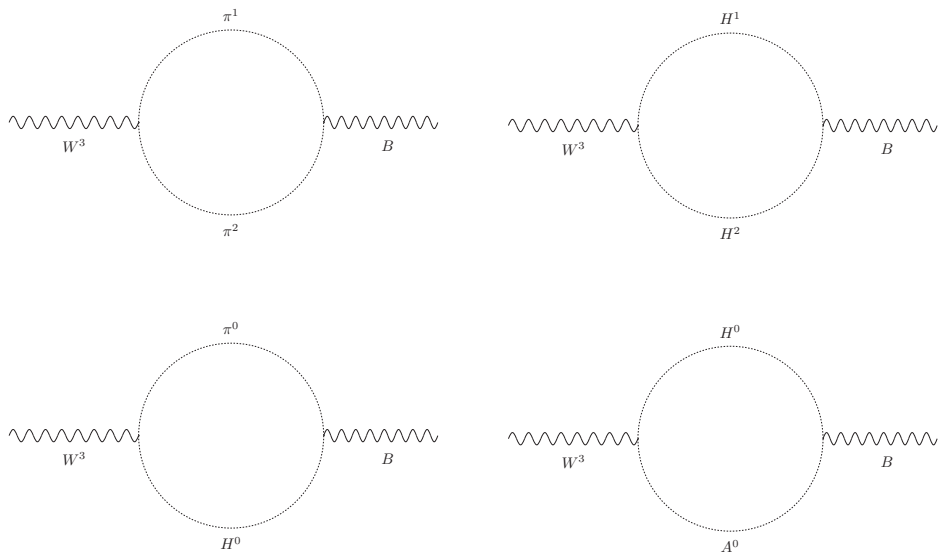}\vspace{-5cm}
\caption{One-loop Feynman diagrams contributing to the $S$ parameter. The
fields $H^1$ and $H^2$ are linear combinations of the charged Higgses $%
H^{\pm}$, similarly to how $W^{\pm}$ gauge bosons are defined in terms of $%
W^1$ and $W^2$.}
\label{diagS1}
\end{figure}

The Feynman diagrams contributing to the $T$ and $S$ parameters are shown in
Figs. \ref{diagT1} and \ref{diagS1}.

\begin{figure}[tbh]
{\includegraphics[width=0.42\textwidth]{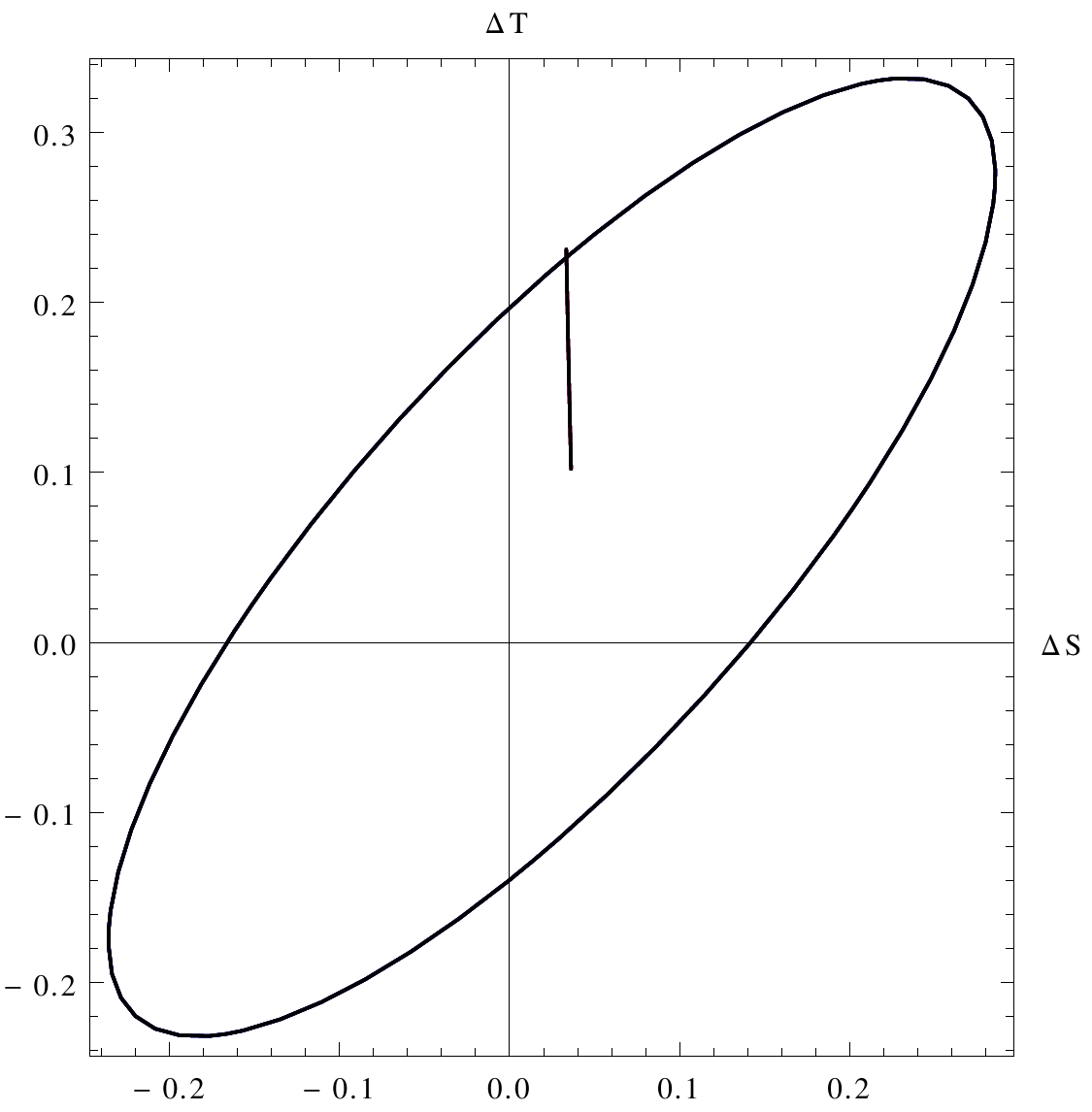}} 
{\includegraphics[width=0.42\textwidth]{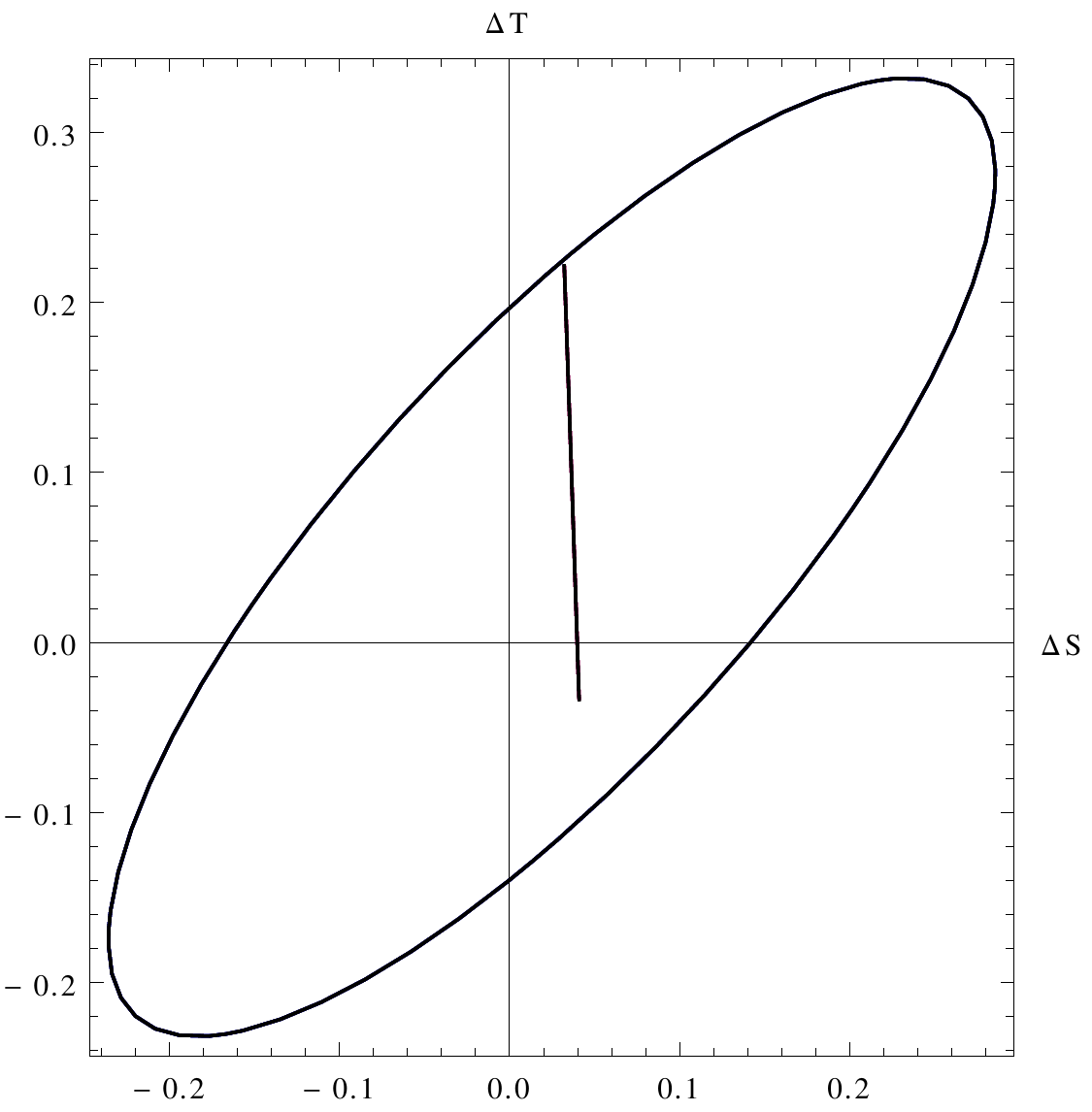}}\newline
{\footnotesize {\ref{TandS1}.a)}}\hspace{5.5cm}{\footnotesize {(\ref{TandS1}%
.b)}} \newline
\caption{The $\Delta S-\Delta T$ plane, where the ellipses contain the
experimentally allowed region at $95\%$ confidence level taken from 
\protect\cite{Baak:2013ppa,Baak:2012kk,Baak:2011ze}. We set $\protect\alpha-%
\protect\beta=\frac{\protect\pi}{5}$. Figures (a) and (b) correspond to $%
m_{A^{0}}=m_{{H}^{0}}=500$ GeV and $m_{H^{0}}=m_{H^{\pm}}=500$ GeV,
respectively. The charged Higgs and CP-odd neutral Higgs boson masses vary
between $550$ GeV $\leq m_{H^{\pm}}\leq$ $580$ GeV (Fig. \protect\ref{TandS1}%
(a), $375$ GeV $\leq m_{A^{0}}\leq$ $495$ GeV (Fig. \protect\ref{TandS1}(b).
The nearly vertical lines going up towards the ellipses correspond to $%
\Delta T$ and $\Delta S$ parameters in our model as masses are varied in the
aforementioned ranges.}
\label{TandS1}
\end{figure}

We split the $T$ and $S$ emphasizing the contributions arising from new
physics as $T=T_{SM}+\Delta T$ and $S=S_{SM}+\Delta S$, where $T_{SM}$ and $%
S_{SM}$ are the SM contributions given by 
\begin{equation}
T_{SM}=-\frac{3}{16\pi \cos ^{2}\theta _{W}}\ln \left( \frac{m_{h}^{2}}{%
m_{W}^{2}}\right) ,  \label{Tt}
\end{equation}

\begin{equation}
S_{SM}=\frac{1}{12\pi }\ln \left( \frac{m_{h}^{2}}{m_{W}^{2}}\right) ,
\end{equation}

while $\Delta T$ and $\Delta S$ contain all the contributions involving in
our model the heavy scalars 
\begin{eqnarray}
\Delta T &\simeq &-\frac{3\cos ^{2}\left( \alpha -\beta \right) }{16\pi \cos
^{2}\theta _{W}}\ln \left( \frac{m_{H^{0}}^{2}}{m_{h}^{2}}\right) +\frac{1}{%
16\pi ^{2}v^{2}\alpha _{EM}(M_{Z})}\left[ m_{H^{\pm }}^{2}-F\left(
m_{A^{0}}^{2},m_{H^{\pm }}^{2}\right) \right]  \notag \\
&&+\frac{\sin ^{2}\left( \alpha -\beta \right) }{16\pi ^{2}v^{2}\alpha
_{EM}(M_{Z})}\left[ F\left( m_{h}^{2},m_{A^{0}}^{2}\right) -F\left(
m_{h}^{2},m_{H^{\pm }}^{2}\right) \right]  \notag \\
&&+\frac{\cos ^{2}\left( \alpha -\beta \right) }{16\pi ^{2}v^{2}\alpha
_{EM}(M_{Z})}\left[ F\left( m_{H^{0}}^{2},m_{A^{0}}^{2}\right) -F\left(
m_{H^{0}}^{2},m_{H^{\pm }}^{2}\right) \right] ,  \label{DeltaT}
\end{eqnarray}

\begin{equation}
\Delta S\simeq \frac{1}{12\pi }\left[ \cos ^{2}\left( \alpha -\beta \right)
\ln \left( \frac{m_{H^{0}}^{2}}{m_{h}^{2}}\right) +\sin ^{2}\left( \alpha
-\beta \right) K\left( m_{h}^{2},m_{A^{0}}^{2},m_{H^{\pm }}^{2}\right) +\cos
^{2}\left( \alpha -\beta \right) K\left(
m_{H^{0}}^{2},m_{A^{0}}^{2},m_{H^{\pm }}^{2}\right) \right] ,  \label{DeltaS}
\end{equation}
where we introduced the functions \cite{Bertolini:1985ia,Hollik:1986gg,Hollik:1987fg,Gunion:1989we,Froggatt:1991qw,Grimus:2008nb,Haber:2010bw,Hernandez:2015rfa}
\begin{equation}
F\left( m_{1}^{2},m_{2}^{2}\right) =\frac{m_{1}^{2}m_{2}^{2}}{%
m_{1}^{2}-m_{2}^{2}}\ln \left( \frac{m_{1}^{2}}{m_{2}^{2}}\right) ,\hspace{%
1.5cm}\hspace{1.5cm}\lim_{m_{2}\rightarrow m_{1}}F\left(
m_{1}^{2},m_{2}^{2}\right) =m_{1}^{2},
\end{equation}%
\begin{eqnarray}
K\left( m_{1}^{2},m_{2}^{2},m_{3}^{2}\right) &=&\frac{1}{\left(
m_{2}^{2}-m_{1}^{2}\right) {}^{3}}\left\{ m_{1}^{4}\left(
3m_{2}^{2}-m_{1}^{2}\right) \ln \left( \frac{m_{1}^{2}}{m_{3}^{2}}\right)
-m_{2}^{4}\left( 3m_{1}^{2}-m_{2}^{2}\right) \ln \left( \frac{m_{2}^{2}}{%
m_{3}^{2}}\right) \right.  \notag \\
&&-\left. \frac{1}{6}\left[ 27m_{1}^{2}m_{2}^{2}\left(
m_{1}^{2}-m_{2}^{2}\right) +5\left( m_{2}^{6}-m_{1}^{6}\right) \right]
\right\},
\end{eqnarray}%
with the properties 
\begin{eqnarray}
\lim_{m_{1}\rightarrow m_{2}}K(m_{1}^{2},m_{2}^{2},m_{3}^{2})
&=&K_{1}(m_{2}^{2},m_{3}^{2})=\ln \left( \frac{m_{2}^{2}}{m_{3}^{2}}\right) ,
\notag \\
\lim_{m_{2}\rightarrow m_{3}}K(m_{1}^{2},m_{2}^{2},m_{3}^{2})
&=&K_{2}(m_{1}^{2},m_{3}^{2})=\frac{%
-5m_{1}^{6}+27m_{1}^{4}m_{3}^{2}-27m_{1}^{2}m_{3}^{4}+6\left(
m_{1}^{6}-3m_{1}^{4}m_{3}^{2}\right) \ln \left( \frac{m_{1}^{2}}{m_{3}^{2}}%
\right) +5m_{3}^{6}}{6\left( m_{1}^{2}-m_{3}^{2}\right) ^{3}},  \notag \\
\lim_{m_{1}\rightarrow m_{3}}K(m_{1}^{2},m_{2}^{2},m_{3}^{2})
&=&K_{2}(m_{2}^{2},m_{3}^{2}).
\end{eqnarray}
The experimental results on $T$ and $S$ restrict $\Delta T$ and $\Delta S$
to lie inside a region in the $\Delta S-\Delta T$ plane. At the $95\%$
confidence level, these are the elliptic contours shown in Fig. \ref{TandS1}%
. The origin $\Delta S=\Delta T=0$ is the SM value with $m_{h}=125.5$~GeV
and $m_{t}=176$~GeV. We analyze the $T$ and $S$ parameter constraints on our
model by considering two benchmark scenarios, in both keeping $\alpha-\beta=%
\frac{\pi}{5}$. In the first scenario we assume that the CP-even and CP-odd
neutral Higgs bosons have degenerate masses of $500$ GeV, below which the
LHC has not detected any scalars beyond the SM-like state. In this first
scenario, we find that the $T$ and $S$ parameters constrain the charged
Higgs masses to the range $550$ GeV $\leq m_{H^{\pm}}\leq$ $580$ GeV, which
is consistent with the lower bound $m_{H^{\pm}} \gtrsim 500$ GeV obtained
from $b\to s\gamma$ constraints \cite{Misiak:2015xwa}. In the second
scenario, we assume that the charged Higgses and CP-even neutral Higgses
have degenerate masses of $500$ GeV. In this second scenario, the $T$ and $S$
parameter constraints are fulfilled if the CP-odd neutral Higgs boson mass
is in the range $375$ GeV $\leq m_{A^{0}}\leq $ $495$ GeV.

\section{Conclusions \label{Conclusions}}

We have constructed a viable 2-Higgs doublet extension of the Standard Model
which features additionally an $S_3$ flavor symmetry and extra scalars that
break $S_3$. This leads to textures for fermion masses, and consists in an
existence proof of models leading to the quark texture in \cite{Hernandez:2014zsa}. Overall, the model can fit the observed masses, CKM and
PMNS mixing angles very well. The model has in total seventeen effective free parameters, which are fitted to reproduce the experimental values of eighteen observables in the quark and lepton sectors, i.e., nine charged fermion masses, two neutrino mass squared splittings, three lepton mixing parameters, three quark mixing angles and one CP violating phase of the CKM quark mixing matrix. The model predicts one massless neutrino for both normal and inverted hierarchies in the active neutrino mass spectrum as well as an effective Majorana neutrino mass, relevant for neutrinoless double beta decay, with values $m_{\beta \beta }=$ 4 meV and 50 meV, for the normal and the inverted neutrino spectrum, respectively. In the latter case our prediction is within the
declared reach of the next generation bolometric CUORE experiment \cite{Alessandria:2011rc} or, more realistically, of the next-to-next generation tonne-scale $0\nu\beta\beta$-decay experiments. The sum of the light active neutrino masses in our model is $59$ meV and $0.1$ eV for the normal and the inverted neutrino spectrum, respectively, which is consistent with the cosmological bound $\sum^3_{k=1}m_{\nu_k}<0.23$ eV. The additional scalars mediate flavor changing
neutral current processes, but due to the specific shape of the Yukawa
couplings dictated by the flavor symmetry these processes occur only in the
up-type quark sector. In the scalar sector the enlarged field content of the
model leads to constraints from both rare top decays and from a $h \to \gamma \gamma$ rate that can be
distinguished from the SM prediction. Among rare top decays, $t \to c h$ is particularly promising as its branching ratio can reach $%
\mathcal{O}(0.01\%)$ in our model. With respect to the $h \to \gamma \gamma$, we find that it depends only
slightly on the mass of the charged Higgs and the dependence on the quartic
scalar couplings is negligible, but the dominant top quark and vector boson
contributions are modified in our model and allow us to place constraints on
the hierarchy of the $SU(2)$ doublet VEVs ($\beta$) and the mixing of their
CP-even mass eigenstates ($\alpha$) that are much stronger than those
obtained from the up-type quark flavor changing processes. We also showed
for a few benchmark scenarios that our model is compatible with the present
bounds for the oblique parameters $T$ and $S$.

\section*{Acknowledgments}

This project has received funding from the European Union's Seventh
Framework Programme for research, technological development and
demonstration under grant agreement no PIEF-GA-2012-327195 SIFT. A.E.C.H
thanks Southampton University for hospitality where part of this work was
done. A.E.C.H was supported by Fondecyt (Chile), Grant No. 11130115 and by
DGIP internal Grant No. 111458.

\appendix

\section{The product rules for $S_{3}$.}

The $S_{3}$ group has three irreducible representations: $\mathbf{1}$, $%
\mathbf{1}^{\prime }$ and $\mathbf{2}$. Denoting the basis vectors for two $%
S_{3}$ doublets as $\left( x_{1},x_{2}\right) ^{T}$\ and $\left(
y_{1},y_{2}\right) ^{T}$ and $y^{\prime }$ a non trivial $S_{3}$ singlet,
the $S_{3}$ multiplication rules are \cite{Ishimori:2010au}:

\begin{equation}
\left( 
\begin{array}{c}
x_{1} \\ 
x_{2}%
\end{array}%
\right) _{\mathbf{2}}\otimes \left( 
\begin{array}{c}
y_{1} \\ 
y_{2}%
\end{array}%
\right) _{\mathbf{2}}=\left( x_{1}y_{1}+x_{2}y_{2}\right) _{\mathbf{1}%
}+\left( x_{1}y_{2}-x_{2}y_{1}\right) _{\mathbf{1}^{\prime }}+\left( 
\begin{array}{c}
x_{2}y_{2}-x_{1}y_{1} \\ 
x_{1}y_{2}+x_{2}y_{1}%
\end{array}%
\right) _{\mathbf{2}},  \label{6}
\end{equation}%
\begin{equation}
\left( 
\begin{array}{c}
x_{1} \\ 
x_{2}%
\end{array}%
\right) _{\mathbf{2}}\otimes \left( y^{\prime }\right) _{\mathbf{1}^{\prime
}}=\left( 
\begin{array}{c}
-x_{2}y^{\prime } \\ 
x_{1}y^{\prime }%
\end{array}%
\right) _{\mathbf{2}},\hspace{1cm}\hspace{1cm}\left( x^{\prime }\right) _{%
\mathbf{1}^{\prime }}\otimes \left( y^{\prime }\right) _{\mathbf{1}^{\prime
}}=\left( x^{\prime }y^{\prime }\right) _{\mathbf{1}}.  \label{7}
\end{equation}

\section{Decoupling and $S_3$ VEVs \label{S3VEV}}

We assume that all SM singlet scalars acquire VEVs much larger than the
electroweak symmetry breaking scale. This implies that the mixing angle
between the scalar singlets and the $SU(2)$ doublet scalars is strongly
suppressed since it is of the order of $\frac{v_{1,2}}{\lambda\Lambda}$, as
follows from the method of recursive expansion of Refs. \cite{Grimus:2000vj,Alvarado:2012xi,Hernandez:2013mcf}. Consequently, the mixing
between these scalar singlets and the SM Higgs doublets can be neglected. We
also checked numerically that the masses of the low-energy scalars are
nearly unaffected by SM singlet VEVs of $\mathcal{O}(500\,\text{GeV})$ and
higher.

For simplicity we assume a CP invariant scalar potential with only real
couplings as done in Refs. \cite{Hernandez:2014lpa,Hernandez:2014vta,Campos:2014zaa,Hernandez:2015cra}. In
the regime where the VEVs decouple, and also because the $1^{\prime }$
scalar $\zeta $ is charged under $Z_{3}^{\prime }$, the relevant terms for
determining the direction of the $\xi $ VEV in $S_{3}$ are 
\begin{equation}
V(\xi )=-\mu _{\xi }^{2}(\xi \xi )_{1}+\gamma _{\xi ,3}(\xi \xi )_{2}\xi
+\kappa _{\xi ,1}(\xi \xi )_{1}(\xi \xi )_{1}+\kappa _{\xi ,2}(\xi \xi
)_{2}(\xi \xi )_{2}+\kappa _{\xi ,3}\left[ \left( \xi \xi \right) _{\mathbf{2%
}}\xi \right] _{\mathbf{2}}\xi ,  \label{Vxi}
\end{equation}

From the minimization conditions of the high-energy scalar potential, we
find the following relations:

\begin{eqnarray}
\frac{\partial \left\langle V\right\rangle }{\partial \text{$v_{\xi _{1}}$}}
&=&2v_{\xi _{1}}\left[ \text{$\mu _{\xi }^{2}+2\left( \kappa _{\xi
,1}+\kappa _{\xi ,2}+\kappa _{\xi ,3}\right) \left( v_{\xi _{1}}^{2}+v_{\xi
_{2}}^{2}\right) $}\right] +3\gamma _{\xi ,3}\left( v_{\xi _{2}}^{2}-v_{\xi
_{1}}^{2}\right) =0  \notag \\
\frac{\partial \left\langle V\right\rangle }{\partial \text{$v_{\xi _{2}}$}}
&=&2v_{\xi _{2}}\left\{ \left[ \text{$\mu _{\xi }^{2}+2\left( \kappa _{\xi
,1}+\kappa _{\xi ,2}+\kappa _{\xi ,3}\right) \left( v_{\xi _{1}}^{2}+v_{\xi
_{2}}^{2}\right) $}\right] +3\gamma _{\xi ,3}v_{\xi _{1}}\right\} =0,
\label{DV}
\end{eqnarray}

Then, from an analysis of the minimization equations given by Eq. (\ref{DV}%
), we obtain for a large range of the parameter space the following VEV
direction for $\xi$: 
\begin{equation}
\left\langle \xi \right\rangle =v_{\xi }\left( 1,0\right).  \label{VEVxi}
\end{equation}

From the expressions given in Eq. (\ref{DV}), and using the vacuum
configuration for the $S_{3}$ scalar doublets given in Eq. (\ref{VEV}), we
find the relation between the parameters and the magnitude of the VEV: 
\begin{equation}
\mu _{\xi }^{2}=-\frac{v_{\xi }}{2}\left[ 3\gamma _{\xi ,3}+4\left( \kappa
_{\xi ,1}+\kappa _{\xi ,2}+\kappa _{\xi ,3}\right) v_{\xi }\right] ,
\label{mu}
\end{equation}

These results show that the VEV direction for the $S_{3}$ doublet $\xi $ in
Eq. (\ref{VEV}) is consistent with a global minimum of the scalar potential
of our model.

\end{document}